\documentclass[sigplan,10pt]{acmart}

\usepackage{booktabs}

\usepackage{amssymb}
\usepackage{bbding}

\usepackage{pifont}
\usepackage{graphicx} % Required for inserting images
\usepackage{amsmath}
\usepackage{multirow}
\usepackage{subfigure}
\usepackage{makecell}
\usepackage[ruled,linesnumbered, noend]{algorithm2e}
\usepackage{enumitem}
\usepackage{xcolor}
\usepackage{lipsum}
\usepackage{comment}
\usepackage{hyperref}

\AtBeginDocument{%
  \providecommand\BibTeX{{%
    \normalfont B\kern-0.5em{\scshape i\kern-0.25em b}\kern-0.8em\TeX}}}

\copyrightyear{2027}
\acmYear{2027}
\setcopyright{cc}
\setcctype{by}
\acmConference[EuroSys '27]{22nd European Conference on Computer Systems}{April 19--23, 2027}{Rabat, Morocco}
\acmBooktitle{22nd European Conference on Computer Systems (EuroSys '27), April 19--23, 2027, Rabat, Morocco}
\acmDOI{10.1145/3842654.3848526}
\acmISBN{979-8-4007-2971-3/2027/04}

\title{Dynamic Flow, Static Graph: KV Cache Reuse for Efficient LLM Serving on Mobile NPUs}

\author{Zhengxiang Huang}
\orcid{0009-0000-1536-1616}
\affiliation{%
  \institution{Shanghai Jiao Tong University}
  \city{Shanghai}
  \country{China}
}
\author{Shengheng Chen}
\orcid{0009-0003-7249-947X}
\affiliation{%
  \institution{Shanghai Jiao Tong University}
  \city{Shanghai}
  \country{China}
}
\author{Chaoyue Niu}
\authornote{Chaoyue Niu is the corresponding author (rvince@sjtu.edu.cn).}
\orcid{0000-0002-1650-4233}
\affiliation{%
  \institution{Shanghai Jiao Tong University}
  \city{Shanghai}
  \country{China}
}
\author{Yujie Sun}
\orcid{0009-0000-3314-5159}
\affiliation{%
  \institution{Shanghai Jiao Tong University}
  \city{Shanghai}
  \country{China}
}
\author{Zhaode Wang}
\orcid{0009-0009-0498-7491}
\affiliation{%
  \institution{Alibaba Group}
  \city{Hangzhou}
  \country{China}
}
\author{Zeyu Zhao}
\orcid{0009-0007-5768-166X}
\affiliation{%
  \institution{Shanghai Jiao Tong University}
  \city{Shanghai}
  \country{China}
}
\author{Chengfei Lv}
\orcid{0009-0004-6618-521X}
\affiliation{
  \institution{Alibaba Group}
  \city{Hangzhou}
  \state{Zhejiang}
  \country{China}
}
\author{Fan Wu}
\orcid{0000-0003-0965-9058}
\affiliation{%
  \institution{Shanghai Jiao Tong University}
  \city{Shanghai}
  \country{China}
}
\author{Guihai Chen}
\orcid{0000-0002-6934-1685}
\affiliation{%
  \institution{Shanghai Jiao Tong University}
  \city{Shanghai}
  \country{China}
}
\renewcommand{\shortauthors}{Zhengxiang Huang et al.}

\begin{document}

\begin{abstract}
On-device large language model (LLM) serving is a cornerstone of local-first personal intelligence, offering users data sovereignty, strong privacy guarantees, and freedom from cloud API latency and cost. Although KV caching is widely used to reduce latency in long-context inference, existing designs were primarily optimized for cloud GPUs with dynamic execution environments and abundant memory bandwidth. These architectural assumptions do not hold on mobile NPUs, where computation graphs must be statically compiled and both memory capacity and I/O bandwidth are severely constrained. In this work, we present a compute-storage co-design for mobile-centric prefix and non-prefix KV reuse. We first propose an intra-graph mechanism that maps selective KV recomputation onto static NPU graphs, reconciling algorithmic dynamicity with NPU staticity. We further develop an inter-graph scheduler to optimize chunk merging and minimize padding with dynamic programming. To address mobile bandwidth limitations, we introduce a hierarchical KV manager featuring a tree-hash-semantic hybrid structure, along with cost-aware prefetching and eviction policies. We also build a two-dimensional pipeline that overlaps KV loading, rerotation, and storage with NPU execution, hiding data-movement latency. Experiments across representative on-device workloads and LLMs show that our design reduces time-to-first-token (TTFT) by 40–60\% compared with no reuse and prefix-only caching.

%On-device large language models (LLMs) are increasingly promising for personalized and privacy-sensitive mobile applications such as personal chatbots, retrieval-augmented generation (RAG), and agents, where long contexts (e.g., chat histories, local documents, and agent skills) are essential. However, such long contexts significantly increase prefill latency, especially on resource-constrained mobile NPUs.  Existing KV prefix and non-prefix reuse systems are designed for cloud GPU serving, yet challenged by mobile NPU offline-compiled static execution pipelines and resource-constrained memory hierarch. On the other hand, existing mobile LLM inference engines collectively lack support for KV reuse mechanisms.  In this paper, we present the first on-device NPU-oriented KV reuse and cache management system supporting both prefix reuse and non-prefix reuse and selective recomputation for long-context LLM serving.  For efficient NPU computation, our design introduces an optimized statically compiled mixed-precision NPU recomputation graph for non-prefix reuse and a CPU-NPU overlapped graph execution pipeline.  For efficient caching and reuse, we further develop a hierarchical KV storing system spanning CPU-NPU memory and flash storage with spatial-temporal-semantic locality aware eviction and reuse policies. Evaluations on 3 smartphones across multiple long-context workloads on Qwen3/Llama3.2 models show that our system achieves 1.4--3$\times$ prefill speedup over ExecuTorch, substantially reducing TTFT for practical on-device LLM applications.
\end{abstract}

\begin{CCSXML}
<ccs2012>
<concept>
<concept_id>10003120.10003138.10003141</concept_id>
<concept_desc>Human-centered computing~Ubiquitous and mobile devices</concept_desc>
<concept_significance>500</concept_significance>
</concept>
<concept>
<concept_id>10010147.10010178.10010179</concept_id>
<concept_desc>Computing methodologies~Natural language processing</concept_desc>
<concept_significance>300</concept_significance>
</concept>
</ccs2012>
\end{CCSXML}

\ccsdesc[500]{Human-centered computing~Ubiquitous and mobile devices}
\ccsdesc[300]{Computing methodologies~Natural language processing}

\keywords{On-Device LLM Inference, Mobile NPU, KV Cache Reuse}

\maketitle

\section{Introduction}
\label{introduction}

The rapid advancement of large language models (LLMs) is driving a fundamental shift in how intelligence is delivered to end users. Rather than relying solely on cloud-centric intelligence, driven by the imperatives of low cost, user privacy, offline availability, and low-latency interaction, leading technology companies are increasingly moving toward a local-first, user-centric intelligence paradigm that enables context-aware, continuously available assistance grounded in sensitive personal data. For example, Google Gemini emphasizes personalized intelligence experiences \cite{google2026gemini}, while Google AI Edge Gallery~\cite{GoogleAI} demonstrates practical deployment of generative models directly on smartphones; Apple Intelligence~\cite{Apple-Intelligence} adopts a hybrid device--cloud collaborative architecture~\cite{Apple-Intelligence-Device-Cloud}, where an on-device 3B-scale LLM is responsible for privacy-sensitive and personalized tasks. Recent systems such as ClawMobile~\cite{ClawMobile} and OmniInfer~\cite{OmniInfer} further demonstrate the growing feasibility of deploying autonomous LLM agents directly on mobile devices.

%On-device LLM inference forms the foundation of such local-first intelligence systems.
%intelligence directly onto personal devices—smartphones, tablets, and laptops—
%Recent advances in On-device LLM service have driven growing interest in local-first personal intelligence, motivated by stronger privacy guarantees, personalized user experiences, and independence from cloud latency and serving cost. 
%Major industry efforts have increasingly shifted toward on-device such user-centric local services. 

%On-device LLM serving is the system foundation of local-first intelligence stack, but face challenges in processing \emph{long-context} and \emph{repeated-context} workloads, which are central to personal intelligence. Consider representative use cases: a long-running conversational assistant whose prompt grows progressively with chat history; a document QA assistant that repeatedly retrieves the same set of user documents; a personal agent that loads shared skill descriptions or tool definitions for every invoked task. In all these scenarios, substantial portions of each prompt, ranging from system prompts and retrieved document chunks to reusable skill instructions, recur across successive requests, creating substantial redundancy in the transformer prefill phase. Figure~\ref{fig:et_latency} shows that the prefill latency, i.e., time-to-first-token (TTFT), can increase to several seconds when the prompt length exceeds 2k tokens and grows linearly as the prompts lengthen, severely degrading user experience.

\begin{figure}[!t]
    \centering
    \subfigure[On-device prefill latency under different prefill length on Xiaomi 15 Pro, inference backed by ExecuTorch, using mobile NPU, 2-5$\times$ slower than cloud. (Qwen3-1.7B, batch size = 1) \label{fig:et_latency}]{\includegraphics[width=0.492\linewidth]{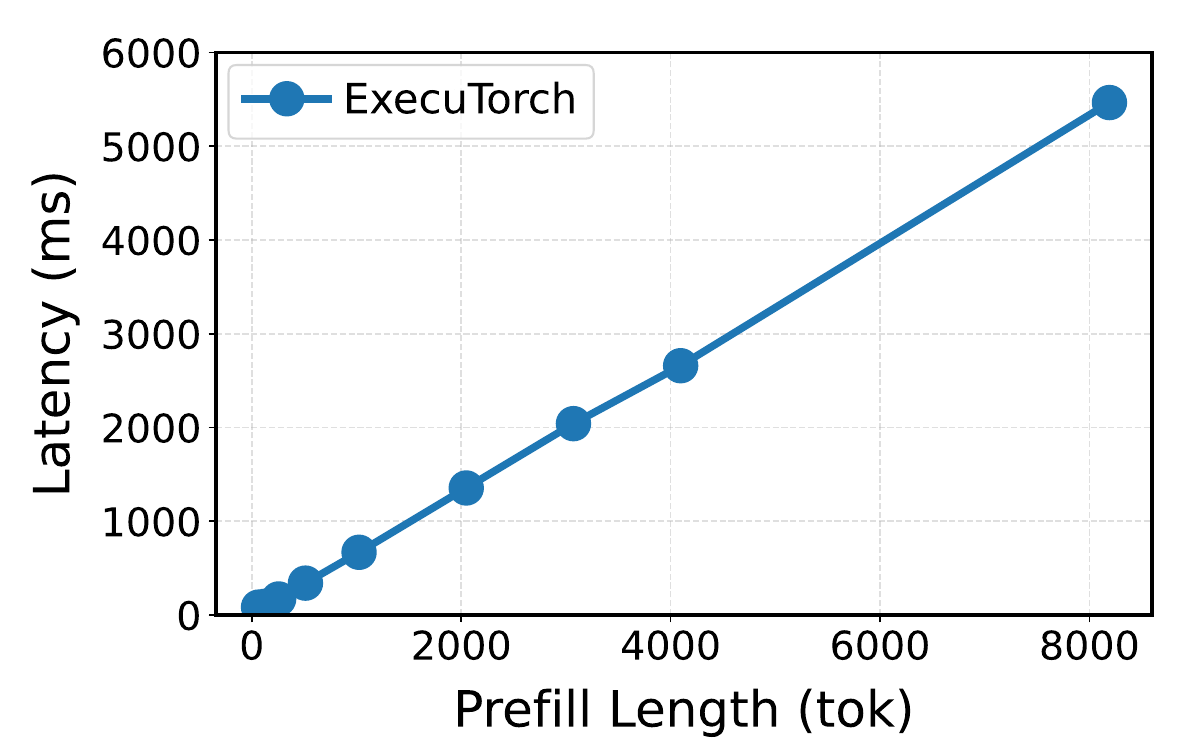}}
    \hfill
    \subfigure[Cloud prefill latency under different prefill length on RTX 5090. Reduction \textcircled{1} comes from prefix caching, while \textcircled{2} comes from non-prefix reuse. (Qwen3-1.7B, batch size = 16)\label{fig:vllm_latency}]{\includegraphics[width=0.492\linewidth]{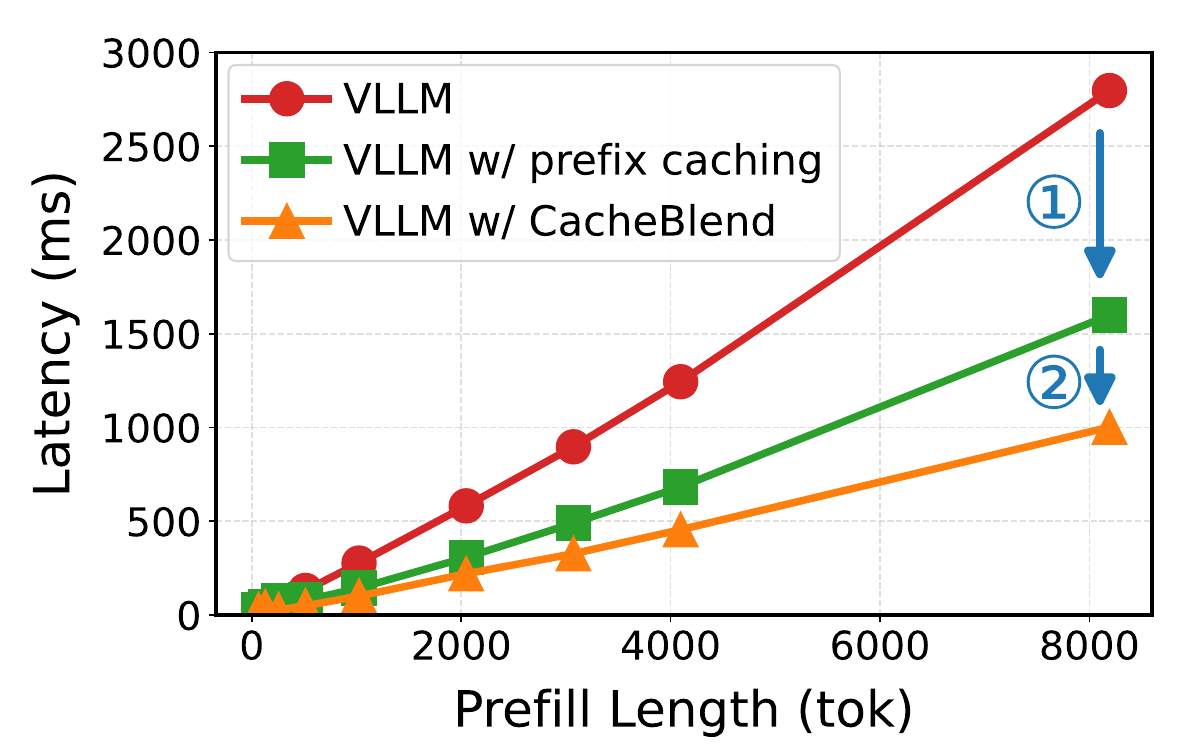}}
    \caption{Long-context prefill latency on mobile devices and KV cache reuse for latency reduction in cloud LLM serving.}
    \label{fig:problem-formulation}
\end{figure}

\begin{figure*}[!ht]
    \centering
    \includegraphics[width=0.92\linewidth]{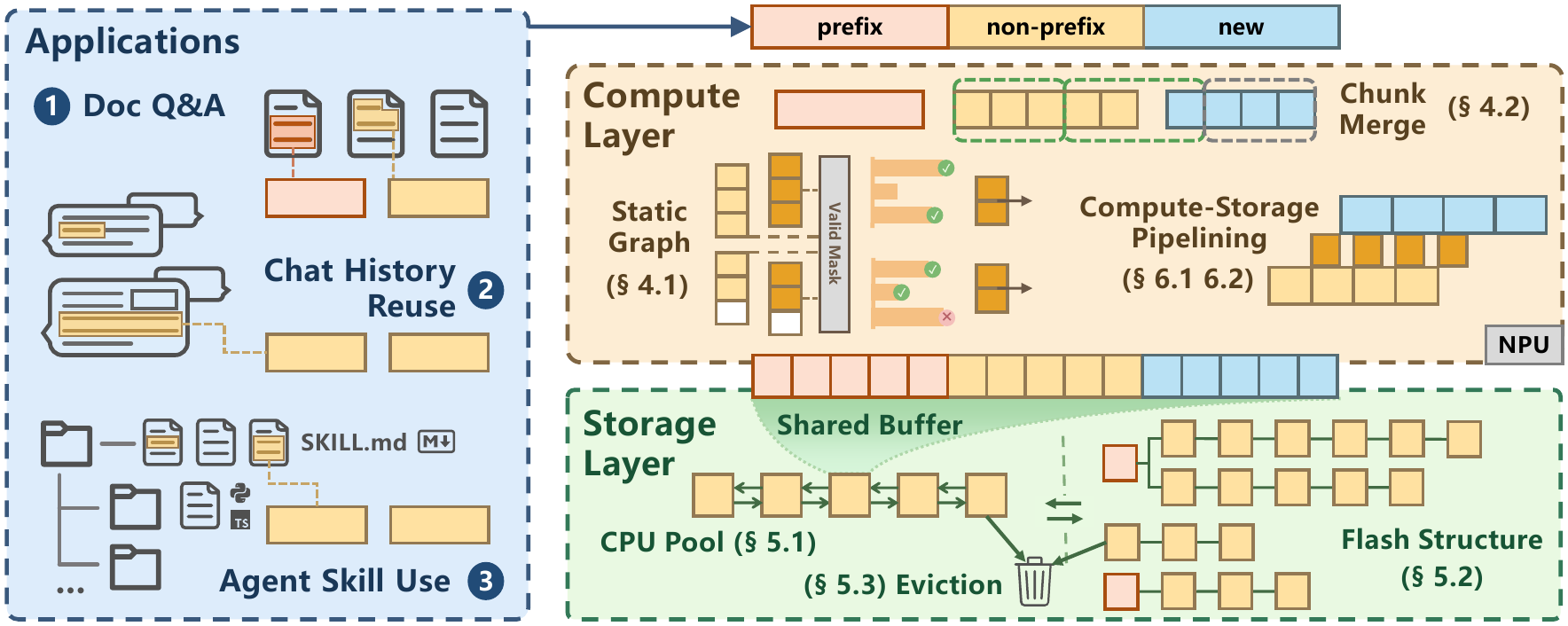}
    % \fbox{
    % \begin{minipage}[c][0.3\textheight][c]{\linewidth}
    %         \centering
    %         \Large Placeholder for Overview Figure
    %     \end{minipage}
    % }
    \caption{System overview of our on-device KV reuse design for long-context mobile LLM serving. The system jointly optimizes compute and storage for both prefix and non-prefix KV reuse, converting dynamic selective recomputation into efficient static NPU execution while enabling hierarchical KV management and asynchronous compute–storage pipeline overlap.}
    %\caption{System overview of our design. Targeting acceleration for long-context mobile LLM applications, our system introduces coordinated KV cache reuse optimizations across both compute and storage layers, and pipelines their interactions to form an efficient on-device NPU KV reuse system.}
    \label{fig:overview}
\end{figure*}

On-device LLM serving forms the system foundation of the emerging user-centric, local-first intelligence stack. 
Like cloud services for personalized intelligence, mobile LLM services face significant challenges in handling \emph{long-context} and \emph{repeated-context} workloads.
Such workload characteristics originate from how these applications construct their prompts, and are therefore shared by cloud and device LLM services alike~\cite{Apple-Intelligence, ClawMobile, RAGCache, LMCache}.
Consider several representative scenarios: a long-running conversational assistant whose prompt grows progressively with chat history; a document QA assistant that repeatedly retrieves the same set of user documents; a personal agent that loads shared skill descriptions or tool definitions for every invoked task.
In each case, context grows long through accumulation or retrieval and becomes reusable because the same content recurs across requests.
Large portions of the prompt, including system prompts, retrieved document chunks, and reusable skill instructions, are repeatedly reused across successive requests, introducing substantial redundancy during the transformer prefill phase.
As illustrated in Figure~\ref{fig:et_latency}, prefill latency, i.e., time-to-first-token (TTFT), increases rapidly with prompt length, reaching several seconds once prompts exceed 2k tokens and continuing to grow nearly linearly thereafter, severely degrading the interactive user experience. 
While the \emph{long and repeated context} workload characteristic is dictated by the application layer rather than the hardware beneath, its cost is far higher on mobile. At the same prompt length, mobile NPUs are 2--5$\times$ slower than cloud GPUs (Figure~\ref{fig:vllm_latency}).

%A mobile assistant rarely answers isolated, stateless prompts. Instead, it repeatedly reasons over long chat histories, personal documents, retrieved knowledge chunks, and reusable tool or skill descriptions. Across such requests, much of the prompt content is repeated exactly or near-exactly, creating substantial redundancy in the transformer prefill phase.

%To improve personalization and service quality, these systems increasingly leverage rich local contexts, including long-term chat histories, private knowledge bases, local documents and emails, and agent skill instructions. Such contextual information is incorporated into prompts to support personalized generation and long-context reasoning. Yet, long contexts significantly slow down LLM inference, particularly during the prefill phase, where the input prompt is processed to construct the KV cache. 

To avoid repeated prefill computation and reduce TTFT, previously computed Key-Value (KV) tensors for input tokens can be cached and reused when similar context appears again. Existing KV cache reuse designs (e.g., Mooncake~\cite{Mooncake}, RAGCache~\cite{RAGCache}, and LMCache~\cite{LMCache}) mainly target \textit{cloud-based deployment} scenarios, focusing on GPU inference and distributed storage architectures. These techniques have also been integrated into LLM serving engines such as VLLM~\cite{vllm} and SGLang~\cite{SGLANG}. Beyond prefix reuse, CacheBlend~\cite{CacheBlend} further introduces selective KV recomputation to enable reuse of non-prefix KV segments. As shown in Figure~\ref{fig:vllm_latency}, prefix caching and non-prefix KV reuse can reduce cloud serving latency by 2$\times$--3$\times$ on cache hits for batched requests served on an RTX 5090 GPU.

However, KV cache reuse has not been well studied in on-device LLM serving scenarios. In particular, mobile LLM inference engines (e.g., ExecuTorch \cite{ExecuTorch}, MLC-LLM \cite{MLC-LLM}, llama.cpp \cite{llama.cpp}, MediaPipe \cite{MediaPipe}) focus largely on efficient single-request execution through optimized kernels, heterogeneous CPU/GPU/NPU scheduling, and quantization. They typically provide little or no support for persistent KV reuse across requests as well as runtime hierarchical KV cache management, and essentially no support for quality-preserving non-prefix reuse. This gap is increasingly important: the workloads are exactly those that exhibit strong cross-request context repetition. Without reuse, mobile systems repeatedly recompute the same long contexts, wasting time, resources, and energy.

% As is illustrated in Table \ref{tab:system-comparison}, 
%Existing cloud systems fail to adapt to these differences in that their cloud GPU targeting solutions do not fit mobile NPU computation and their storage systems do not fit on-device resource-constrained storage either.
%On the other hand, current on-device  systems, such as MLC-LLM \cite{MLC-LLM}, MediaPipe \cite{MediaPipe}, ExecuTorch \cite{ExecuTorch}, and llama.cpp \cite{llama.cpp}, 
% ORT \cite{ORT},
% mllm \cite{mllm}, and MNN \cite{Walle, MNN-LLM} 
%focus on LLM inference optimization, and do not support persistent and runtime hierarchical KV cache management, let alone advanced prefix or non-prefix caching and reuse. 

%However, existing KV-reuse techniques are designed primarily for cloud hardware and engines, and they do not transfer directly to mobile devices. 

Directly porting cloud-based KV reuse techniques to mobile devices is fundamentally infeasible due to profound architectural differences between cloud and mobile hardware. 
% This challenge stems from a fundamental mismatch between KV reuse algorithms and mobile hardware characteristics. 
First, from a compute perspective, mobile NPUs are architecturally distinct from cloud GPUs. Whereas cloud GPUs expose flexible, programmable tensor cores with SIMT parallelism that naturally accommodate dynamic tensor shapes and irregular computational graphs, mobile NPUs are built around fixed-shape matrix tile units that operate on statically pre-compiled computational graphs. Dynamic execution patterns, such as the variable token selection, input-dependent KV deviation computation, and runtime reuse-pattern variation, introduced by non-prefix selective recomputation, are fundamentally incompatible with this static execution model. Second, from a storage perspective, on-device platforms operate under memory bandwidth and capacity constraints that are orders of magnitude tighter than cloud deployments, creating a severe memory and bandwidth wall. Loading and storing large KV tensors can easily exceed the time saved by bypassing the prefill computation. Without a specialized hierarchical management and prefetching strategy that considers both NPU execution and flash I/O, the overhead of KV movement can negate the benefits of reuse.
These differences are structural rather than transient: mobile SoCs remain constrained by stringent power and thermal budgets, and thus cannot scale compute and memory resources in the same manner as datacenter accelerators.
Consequently, although long-context and context-reuse workloads are shared across device and cloud services, efficiently exploiting such reuse opportunities on mobile devices requires fundamentally novel engine-level designs rather than directly inheriting cloud-oriented mechanisms.

In this work, we build the first system that enables efficient prefix and non-prefix KV reuse for on-device LLM serving on mobile NPUs, as depicted in Figure~\ref{fig:overview}.
Unlike existing cloud designs that target cloud GPU kernels and distributed clusters, our key idea is a compute-storage co-design that adapts dynamic KV reuse algorithms to the static, quantized execution model of mobile NPUs while simultaneously respecting the severe bandwidth and capacity constraints of on-device memory hierarchies. 
On the {\em compute side} (\S~\ref{compute-design}), 
to bridge the gap between dynamic algorithms and static hardware, we first map selective recomputation onto pre-compiled static NPU graphs, using chunk-and–pad, validity masking, and forced final-token selection to preserve irregular reuse semantics under fixed input shapes. At the inter-graph level, we formulate runtime execution as a multi-graph scheduling problem and develop a novel dynamic programming scheduler that merges invocations across a set of pre-compiled static graphs to minimize latency under dynamic input.
On the \emph{storage side} (\S~\ref{storage-design}), to fit on-device budgets while avoiding stalling NPU prefill, we design a multi-tier storage manager spanning the NPU buffer, CPU memory, and flash. It uses a hybrid on-disk data structure combining prefix trees with hash and semantic indices to efficiently locate both prefix and non-prefix reusable chunks. It also takes a lightweight, cost-aware prefetch and eviction policy that incorporates the differential cost of reloading versus recomputing prefix and non-prefix chunks, ensuring hot KV data stays close to the compute units. 
To prevent data movement from dominating latency (\S~\ref{pp}), we jointly schedule storage I/O and CPU-NPU execution based on parallelizability analysis across model shards and input chunks.
Together, these components form a cross-layer co-design with mutually complementary compute and storage decisions, as validated by our ablation study.

We summarize the key contributions as follows:
\begin{itemize}
    \item We identify fundamental system bottlenecks in on-device LLM serving with KV reuse: the architectural mismatch between dynamic KV recomputation and the static compiled graphs required by mobile NPUs, exacerbated by heavy KV cache transfer under tight on-device memory bandwidth and capacity.

    \item We develop an intra-graph construction and a general multi-graph scheduling abstraction for dynamic inputs over pre-compiled static graphs, along with a cross-layer compute--storage design that jointly optimizes KV recomputation, placement, and movement.

    \item 
    We take 3 different workloads, 4 LLMs from Qwen3 and Llama3.2 series, and 3 smartphones for evaluation. Evaluation results reveal that our design delivers 1.5$\times$--5.4$\times$ prefill speedup over original ExecuTorch~\cite{ExecuTorch} with no reuse and 1.3$\times$--2.5$\times$ over prefix caching, cuts TTFT by 40\%–60\%, and avoids the roughly 30\% quality drop of direct full non-prefix reuse.
\end{itemize}

\begin{comment}
\begin{enumerate}
    \item We enable efficient prefix and non-prefix KV reuse on mobile NPUs.
    \item We design an on-device hierarchical KV caching system to supply rapid and stable loading and storing of cached KV tensors.
    \item We build a complete compute--storage overlapped KV reuse pipeline for long-context mobile LLM serving.
\end{enumerate}
\end{comment}

\section{Background}
\label{background}
\subsection{KV Cache Prefix and Non-Prefix Reuse}
\label{bg-reuse}
For transformer-based LLMs, the prefill phase computes per-layer Key-Value (KV) tensors for the input tokens. 
These tensors form the KV cache, which can be reused when the same context reappears, thereby reducing redundant prefill computation and reducing TTFT.

\begin{figure}[h]
    \centering
    \includegraphics[width=0.9\linewidth]{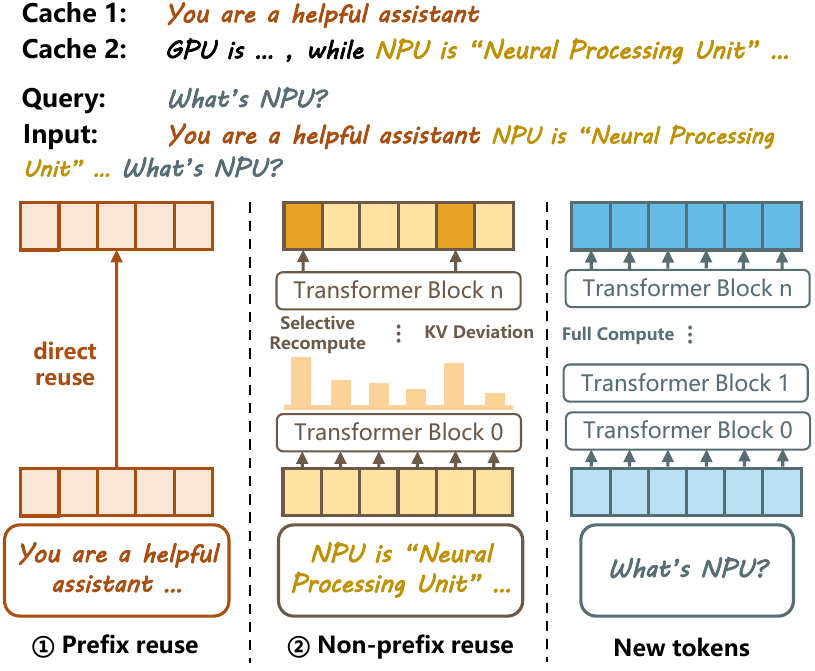}
    %\vspace{-0.3em}
    \caption{Illustration of 3 types of tokens in a prompt: prefix, non-prefix, and new tokens. \textcircled{1} Prefix caching directly reuses KV states for shared prefixes. \textcircled{2} Non-prefix KV chunks recompute only high KV deviated tokens and reuse the rest.}
    \label{fig:reuse-intro}
\end{figure}

For input prompt, tokens can be categorized into 3 types: \emph{prefix reusable}, \emph{non-prefix reusable}, and \emph{new} tokens, as exemplified in Figure \ref{fig:reuse-intro}. 
\textbf{Lossless prefix caching} strategy directly reuses the KV cache of a prompt prefix when a new request shares the same leading tokens as some previously cached prompts (\textcircled{1} in Figure \ref{fig:reuse-intro}).
Because the KV states of a prefix do not depend on subsequent tokens, this reuse is lossless and preserves generation quality, therefore widely adopted in modern cloud-based LLM serving systems, such as VLLM~\cite{vllm}, SGLANG~\cite{SGLANG}, and ChunkAttention~\cite{ChunkAttention}, but its benefit is limited to prefix reusable texts that posit at the beginning of the prompt.

A more flexible but \textbf{lossy non-prefix chunk reuse} strategy targets  reusable non-prefix chunks in the prompts, offering a substantially larger reusing space. 
Direct reusing non-prefix chunks in Prompt Cache ~\cite{PromptCache} may significantly lose accuracy because they cannot recover the true positional dependency and cross-chunk attention of non-prefix chunks. 
Instead, CacheBlend~\cite{CacheBlend}, which recomputes a small set of selected tokens with high KV deviation while reusing the remaining chunk-level KV cache (\textcircled{2} in Figure \ref{fig:reuse-intro}), provides a practical middle ground between exact but restrictive prefix reuse and expensive full recomputation for non-prefix reuse. 
Based on the insight that KV deviation is strongly correlated across layers~\cite{CacheBlend}, the algorithm compares cached KV with fully recomputed KV in the first few layers, selects the top-$K$ most deviated tokens, and selectively recomputes them in the subsequent layers. 
Thus, it reduces the computational FLOPs to approximately the recompute ratio $r$ of full computation and induces only tiny precision loss. 
% Because of its strong precision-efficiency tradeoff, we adopt CacheBlend as the non-prefix reuse algorithm in this work.

\begin{figure}[!h]
    \centering
    \subfigure[Selective KV recompute algorithm workflow with 5 forms of dynamicity.]{\includegraphics[width=0.5\linewidth]{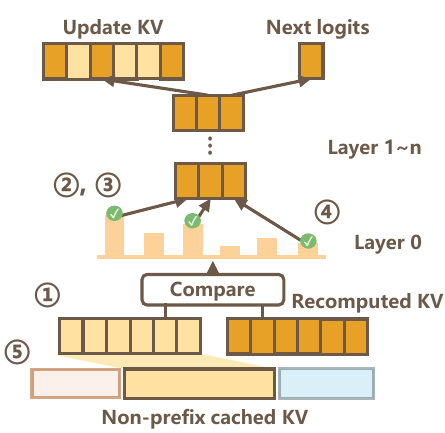}\label{fig:cloud-dynamic-graph}}
    \hfill
    \subfigure[Heatmap of KV deviation $\Delta KV$.]{\includegraphics[width=0.47\linewidth]{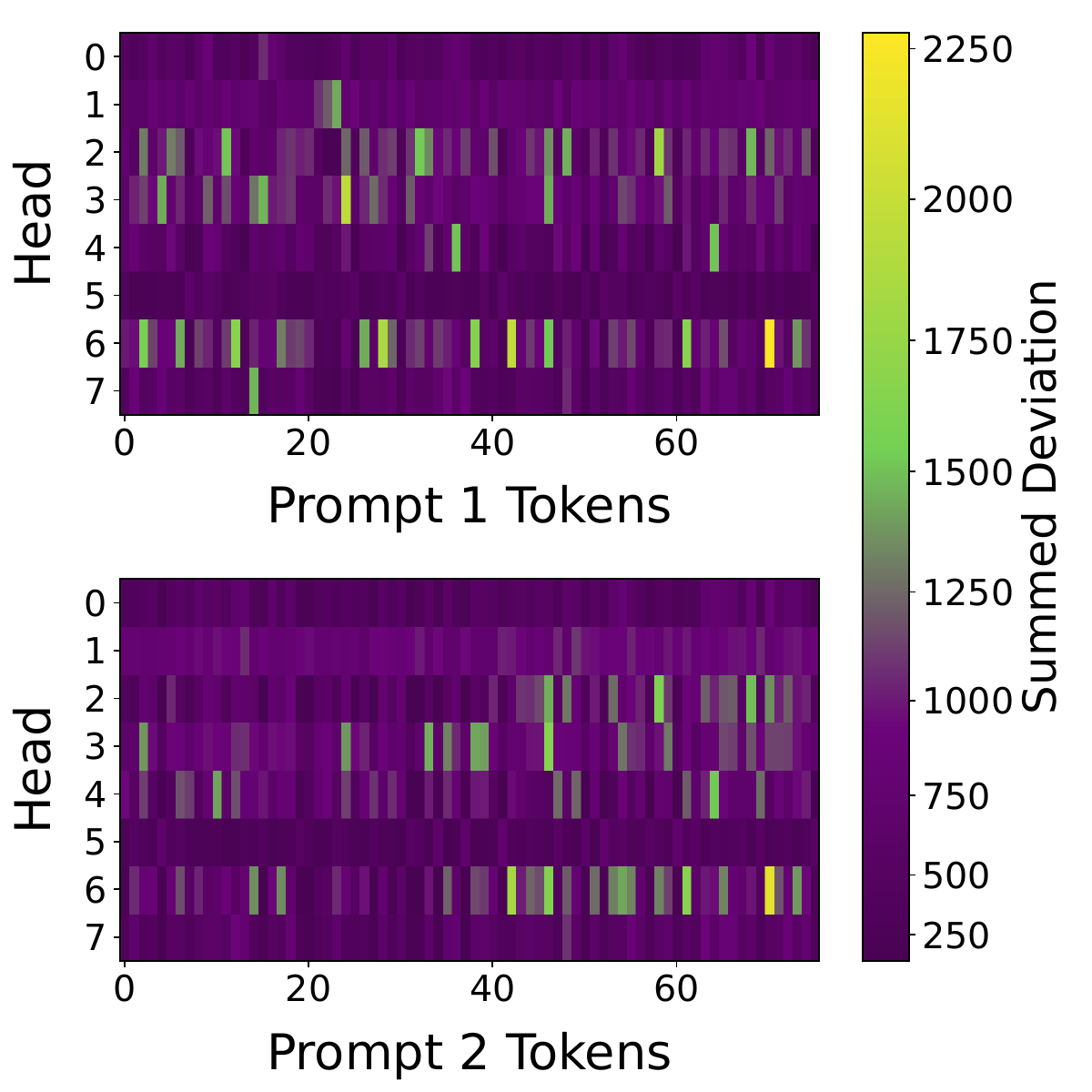}\label{fig:kv-deviation}}
    \caption{Characteristics of non-prefix KV reuse and selective recompute algorithm.}
    \label{fig:compute-challenge}
\end{figure}

As illustrated in Figure~\ref{fig:cloud-dynamic-graph}, the selective non-prefix KV recompute workflow of CacheBlend algorithm features 5 forms of dynamicity in its computational graph.
\textcircled{1} \emph{Input shape dynamicity}, where prompt lengths vary significantly across requests.
\textcircled{2} \emph{Recomputation selection dynamicity}, where the subset of tokens to recompute is determined dynamically according to KV deviation.
\textcircled{3} \emph{Dynamic numerical range}, where the KV comparison operation is highly precision-sensitive. 
Intermediate summed squared error of $i$-th token of the layer 0 is computed as:
\begin{equation}
\Delta KV[i]
=
\left\|
KV_{\mathrm{layer}\,0}[i]
-
\widehat{KV}_{\mathrm{layer}\,0}[i]
\right\|_F^2,
\label{eq:kv-deviation}
\end{equation}
which spans a wide numerical range from 0 to 2000 and exhibits substantially different distributions across prompts as shown in the KV deviation heatmap in Figure~\ref{fig:kv-deviation}.
Such large dynamic ranges make quantization difficult and can lead to overflow or underflow when statically quantized to 8/16-bit integers on mobile NPUs.
\textcircled{4} \emph{Output last token dynamicity}, where the final prompt token must always be selected to produce logits for subsequent autoregressive decoding.
\textcircled{5} \emph{Reuse-pattern dynamicity}, where the token reuse pattern within a prompt is determined at runtime, with prefix-reusable tokens, non-prefix-reusable tokens, and new tokens appearing in arbitrary interleaved combinations and lengths.
% To enable efficient non-prefix KV reuse on mobile NPUs, \textcircled{1}, \textcircled{2}, and \textcircled{3} are addressed in Section~\ref{static-dynamic}, while  \textcircled{4} are addressed in Section~\ref{static-util}.

\subsection{Mobile Hardware Characteristics}
\label{bg-hardware}
\paragraph{Mobile NPU Characteristics}
Mobile NPUs equipped with specialized matrix units can substantially accelerate long-context LLM prefill, often achieving 2--5$\times$ speedup over mobile CPU and GPU~\cite{mllm}. 
These matrix units process fixed-shape tensor tiles streamingly, which strongly favor \textbf{static graphs}, i.e., offline pre-compiled computational graphs with statically determined operator shapes and workflow.
In practice, NPU graph preparation involves graph construction, optimization, and compilation, whose overhead is only acceptable when performed offline. Unlike cloud GPUs, which leverage symmetric tensor cores and SIMT parallelism for flexible programmability and irregular-shape support, mobile NPUs deliberately trade this flexibility for efficiency: regular tensor shapes enable efficient mapping onto matrix units and streamlined hardware scheduling under stringent power and thermal budgets, which is what underlies their performance-per-watt advantage. Consequently, they require specialized techniques to convert dynamic execution patterns into static graphs. 
This execution preference is not vendor-specific, but is shared across major mobile NPU platforms including Qualcomm HTP~\cite{hexagon-htp}, MediaTek APU~\cite{NeuroPilot}, and Apple ANE~\cite{CoreML}, and it persists across the three generations of Hexagon NPUs evaluated in this work.

%Unlike cloud GPUs, which consist of symmetric tensor cores and rely on SIMT parallelism with stronger programmability and irregular-shape handling capability, mobile NPUs require specialized techniques to transform dynamic execution patterns into static graphs.

In addition, mobile NPUs achieve the highest compute and memory efficiency on \textbf{integer GEMM} workloads, such as $W_{\mathrm{INT4}}A_{\mathrm{UINT16}}$~\cite{hexagon-htp,T-MAN}, while floating-point execution is comparatively more expensive. 
Mobile NPUs also favor \textbf{static quantization}, where quantization parameters are calibrated offline and reused during inference. 
Although effective for regular LLM prefill and decode, static quantization is less suitable for selective recomputation, because the compare-and-select stage relies on precision-sensitive KV difference computation whose distributions are highly input-dependent and difficult to calibrate offline.

\begin{figure}[!t]
    \centering
    \includegraphics[width=0.9\linewidth]{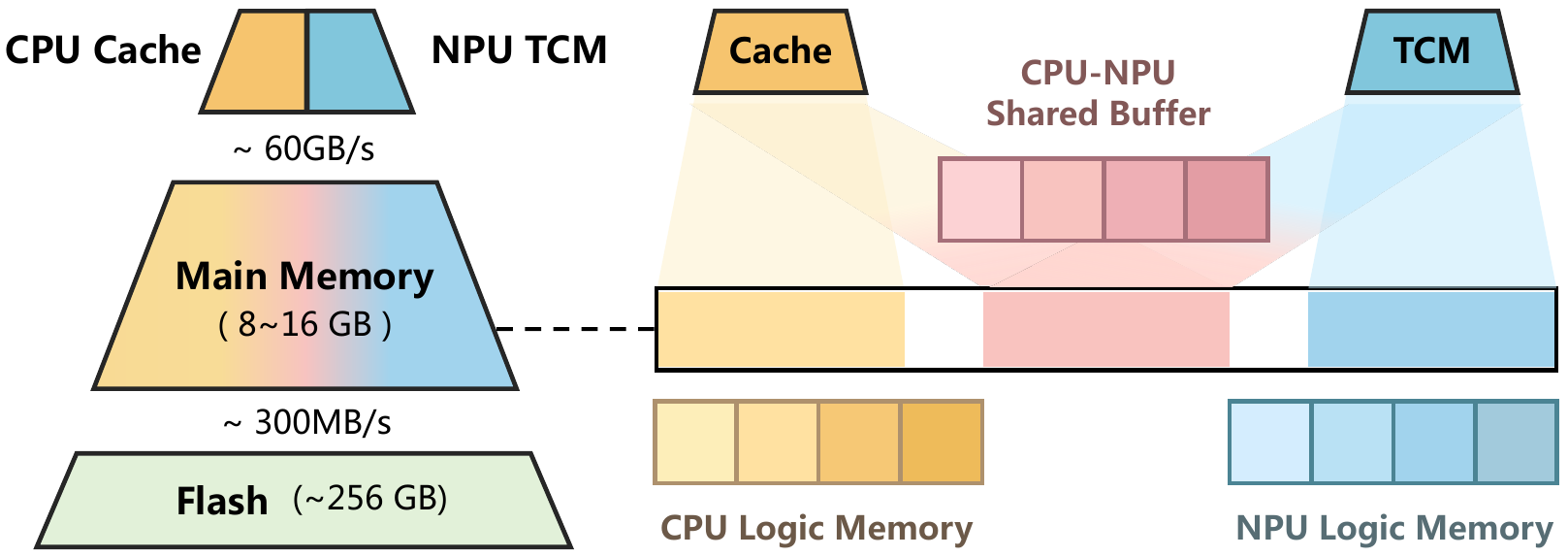}
    \caption{CPU--NPU memory hierarchy of mobile devices.}
    \Description{On-device memory hierarchy}
    \label{fig:bg-mem}
\end{figure}

\paragraph{On-Device Memory Hierarchy}
The on-device memory hierarchy typically follows a three-tier structure, as illustrated in Figure~\ref{fig:bg-mem}: (1) CPU cache and NPU tightly coupled memory (TCM) SRAM at the top; (2) CPU-NPU shared DRAM in the middle; and (3) flash storage at the bottom. While the CPU cache and NPU TCM offer low latency for compute-intensive operations, their capacity is strictly limited (often $<10$\,MB). On modern SoCs like the Qualcomm HTP~\cite{hexagon-htp}, the DRAM layer is physically shared between the CPU and NPU, while logically divided into CPU logic memory, NPU logic memory, and a CPU--NPU shared buffer. At the lowest level, flash storage serves as persistent storage for models and KV data.

\begin{table}[h]
\centering
\caption{Comparison between on-device and cloud memory hierarchies for LLM serving and KV storage.}
\label{tab:mem-compare}
\resizebox{0.98\linewidth}{!}{
\begin{tabular}{llccc}
\toprule
\textbf{Platform} & \textbf{Memory Level} & \textbf{Capacity} & \textbf{Bandwidth} & \textbf{Interconnect} \\
\midrule

\multirow{3}{*}{\textbf{Device}}
& Cache / TCM
& 2--8\,MB
& --
& -- \\

& CPU LPDDR
& 8--16\,GB
& $\sim$60\,GB/s
& On-chip Bus \\

& Flash
& 128\,GB--1\,TB
& $\sim$500\,MB/s
& I/O Bus \\

\midrule

\multirow{4}{*}{\textbf{Cloud}}
& GPU Cache
& $\sim$100\,MB
& --
& -- \\

& GPU HBM
& 24--640\,GB
& 1--3\,TB/s
& NVLink \\

& CPU DRAM
& 0.5--10\,TB
& 50--300\,GB/s
& PCIe \\

& Distributed Storage
& 10--1000\,TB
& 0.5--2\,GB/s
& RDMA \\

\bottomrule
\end{tabular}
}
\end{table}

In contrast to cloud-based LLM serving, on-device platforms operate under substantially tighter memory-capacity and bandwidth constraints. Cloud deployments typically organize memory hierarchically across GPU device memory, CPU host memory, and distributed storage, interconnected through high-bandwidth fabrics such as NVLink, PCIe, and RDMA. Table~\ref{tab:mem-compare} summarizes cloud-based and on-device memory architectural disparities, highlighting the bandwidth wall faced by mobile-class hardware.

%%Compared with cloud-based LLM serving systems, on-device platforms operate under much tighter memory capacity and bandwidth constraints. Cloud systems typically organize memory across GPU device memory, CPU host memory, and distributed storage connected through NVLink, PCIe, and RDMA. Table~\ref{tab:mem-compare} compares the memory characteristics between on-device and cloud-based systems.

\section{New Key Challenges}
\label{challenges}
Given the algorithmic characteristics of prefix and non-prefix KV reuse (\S~\ref{bg-reuse}) and the mobile hardware characteristics (\S~\ref{bg-hardware}), two key challenges emerge from the compute and storage perspectives.

\subsection{Recomputation Dynamicity vs. NPU Staticity}
\label{compute-challenge}

From the compute perspective, the key challenge lies in the mismatch between the highly dynamic execution patterns introduced by selective recomputation for non-prefix reusable tokens and the pre-compiled static execution graphs optimized by mobile NPUs.

%First, non-prefix selective recomputation introduces highly dynamic execution patterns, while mobile NPUs are optimized for pre-compiled static graphs. 
%From the compute perspective, the key challenge lies in the mismatch between the inherent dynamic behavior of selective recomputation algorithm for non-prefix reusable tokens and the static execution flow on mobile NPUs. 

% Selective recomputation dynamically determines tokens with high KV deviation to be recomputed, whereas NPUs rely on pre-compiled graphs with fixed tensor shapes, operator structures, and quantization configurations.

The 5 forms of dynamicity discussed in \S~\ref{bg-reuse} translate into several concrete conflicts with mobile NPU execution. 
Specifically, dynamic tensor shapes (\textcircled{1}) conflict with statically fixed tensor sizes in compiled graphs; dynamic computational flows (\textcircled{2} and \textcircled{4}) conflict with static operator graphs; dynamic numerical ranges (\textcircled{3}) conflict with static quantization parameters; and dynamic reuse patterns (\textcircled{5}) conflict with static graph invocation strategies. 
These conflicts jointly complicate the design of efficient intra-graph construction and quantization schemes, as well as inter-graph scheduling and invocation optimization.

\subsection{Large KV Size vs. Limited Bandwidth and Space} 
\label{storage-challenge}

From the storage perspective, the key challenge arises from the mismatch between the large size of KV tensors and the limited bandwidth and capacity of on-device memory hierarchies, as KV chunks must be loaded from lower memory levels with limited bandwidth and stored under tight memory capacity constraints, making efficient hierarchical KV management difficult.

%From the storage perspective, the key challenge arises from the mismatch between the size of KV tensors and the limited bandwidth and capacity of on-device memory hierarchies.
%Second, KV chunks must be loaded from lower memory levels with limited bandwidth and stored under tight memory capacity constraints, making efficient hierarchical KV management difficult.

Due to constrained I/O bandwidth, loading large KV tensors from flash storage can introduce substantial latency. 
For example, the 8-bit quantized KV cache of Qwen3-1.7B occupies approximately 56\,KB per token, corresponding to 56\,MB for a 1024-token prompt. 
On Xiaomi 15 Pro, loading and mapping such a prompt from flash storage into the memory pool takes approximately 200\,ms, which can diminish the execution advantage of NPUs by causing them to wait if the latency is not carefully hidden.

Limited on-device memory and storage capacity prevent KV tensors from being retained indefinitely. 
As a result, eviction and compaction mechanisms are required to periodically or adaptively reclaim and reorganize storage space. 
These maintenance operations must also be carefully overlapped with execution; otherwise, excessive secondary-storage I/O and fragmentation can significantly degrade performance and expose additional latency to users.

The above new challenges make existing cloud-based KV reuse approaches inapplicable, demanding new compute-storage co-designs for efficient on-device KV reuse.

%These on-device hardware characteristics and constraints make directly applying existing KV reuse approaches fundamentally challenged by, demanding new compute-storage designs for an efficient on-device KV reuse system.

\section{NPU Recompute for Non-Prefix Reuse}
\label{compute-design}
To resolve the conflicts between the dynamicity of non-prefix reuse selective recompute algorithm and the staticity of NPU pre-compiled workflows, we propose intra-graph design that converts dynamic algorithm to static NPU computation graph, searches for the latency-optimal static tensor shape to optimize the graph, and supports precision-sensitive operators with mixed-precision operators to cover dynamic numerical range of intermediate tensors. We further propose inter-graph dynamic programming to optimize dynamic reuse patterns via merging chunks and reducing padding.

%To convert the non-prefix reuse algorithm to static NPU graphs, the dynamicity discussed in \S\ref{bg-reuse} needs to be addressed. Figure~\ref{fig:npu-graph} illustrates the converted static NPU KV recomputation graph.

\subsection{Intra-Graph Design for Dynamic Recomputation}
\label{static-dynamic}
\begin{figure}[t]
    \centering
    \includegraphics[width=0.53\linewidth]{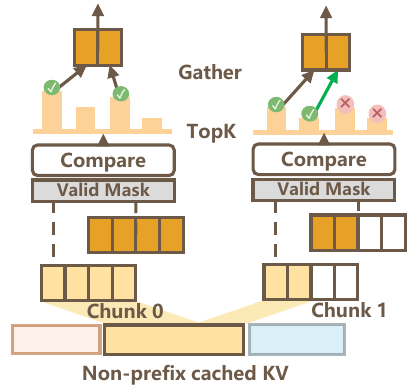}
    \caption{NPU static KV recompute graph on layer 0.}
    \label{fig:npu-graph}
\end{figure}
\paragraph{NPU Static Graph Construction} 
Converting the non-prefix reuse algorithm CacheBlend to static NPU graph requires addressing the dynamicity detailed in \S\ref{bg-reuse}. The resulting static NPU KV recomputation graph is illustrated in Figure~\ref{fig:npu-graph}. The dynamic input shape (dynamicity \textcircled{1}) is addressed by chunking the non-prefix tokens into static size slices, padded if not divisible by the chunk length. A valid mask is applied before KV deviation computation to exclude padded tokens from the Top-K selection process, preventing invalid tokens from being mistakenly selected for recomputation. Additionally, the mask guarantees that the final token is always selected (green arrow in Figure~\ref{fig:npu-graph}) wherever its position in last chunk is (dynamicity \textcircled{4}), ensuring correct generation of the final logits. This \textit{chunk-pad-mask} adapts the static graph to dynamic input and output shapes.

After Top-K token selection, the \textit{scatter-gather} mechanism gathers the recomputed KV of tokens with high KV deviation and scatters them back into the reused KV cache, replacing the corresponding reused KV entries to preserve generation quality, which resolves the dynamicity in token selection and KV update workflow (dynamicity \textcircled{2}).

%for graph execution efficiency. An optimized chunk length can contribute to higher

% \niu{Chunk Length, the prefill graph size $L_P$ and the selective recomputation graph size $L_S$.}

\paragraph{Latency-Optimized Graph Size Configuration} 
The static graph input size is critical to improve NPU graph utilization and execution efficiency, thus maximizing hardware utilization, such as NPU TCM capacity and matrix units (e.g., Qualcomm QNN HMX tiles), while minimizing padding overhead caused by indivisible input shapes. An optimal graph size can be determined by minimizing the expected latency over a representative target workload. In our design, the prefill graph size $L_P$ and the selective recomputation graph size $L_S$ are optimized. Given the latency of one graph invocation, denoted by $\mathrm{Latency}_P(L_P)$ and $\mathrm{Latency}_S(L_S)$, the latency for processing tokens of length $n$ can be expressed as:
\begin{equation}
    \begin{array}{c}
    \mathrm{Latency}_P(n) = \lceil \frac{n}{L_P} \rceil\cdot \mathrm{Latency}_P(L_P),  \\
    \mathrm{Latency}_S(n) = \lceil \frac{n}{L_S} \rceil\cdot \mathrm{Latency}_S(L_S).
    \label{eq:graph-latency}
    \end{array}
\end{equation}
The expected latency of graph $G$ can be computed as:
\begin{equation}
    \mathbb{E}(\mathrm{Latency}_G)=\sum_{n=1}^N \mathrm{freq}(n)\cdot \mathrm{Latency}_G(n),\,G\in \{P,S\}.
\end{equation}
$\mathrm{freq}(n)$ denotes the occurrence frequency of token length $n$ collected from a representative workload.

\begin{figure}[h]
    \centering
    \subfigure[Graph execution per-token latency under different graph sizes.]{\includegraphics[width=0.48\linewidth]{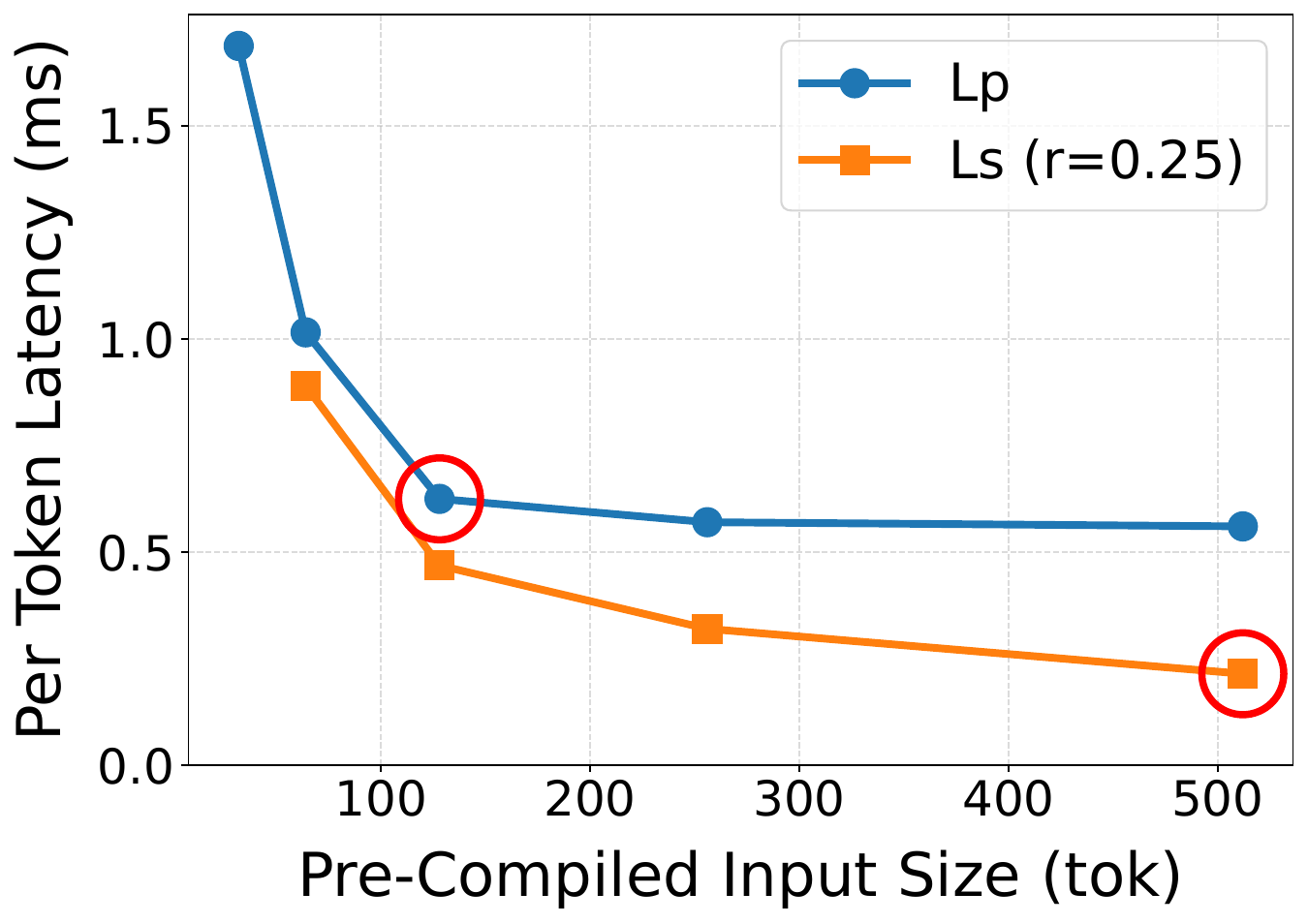}\label{fig:chunk-latency}}
    \hfill
    \subfigure[Non-prefix tokens prefill latency under different input sizes.]{\includegraphics[width=0.48\linewidth]{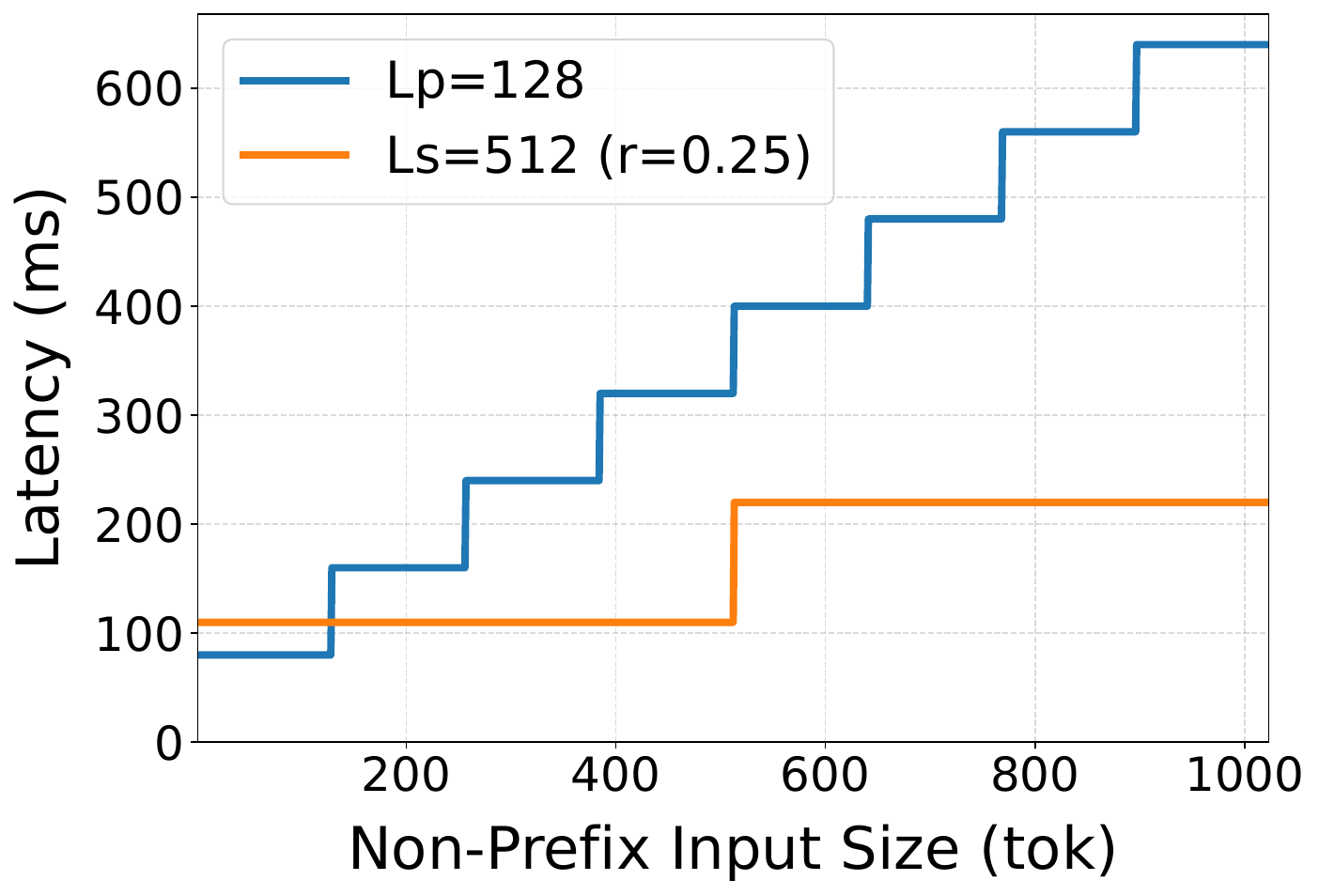}\label{fig:latency-all}}
    \vspace{-0.5em}
    \caption{Latency analysis of static graph input sizes.}
    \vspace{-0.5em}
    \label{fig:optimal-chunking}
\end{figure}

%We constrain graph sizes to multiples of 32, following prior work on Qualcomm HMX tile utilization~\cite{llm-npu}, and select the configuration that minimizes expected latency.
% \niu{In our setting, the optimal graph sizes are $L_P=128$ and $L_S=512$.}

Aligned with Qualcomm HMX tile utilization strategies~\cite{llm-npu}, we restrict graph sizes to multiples of 32 and select the configuration that yields the lowest expected latency. As shown in Figure~\ref{fig:chunk-latency}, per-token latency is initially high for small input lengths due to poor hardware utilization and then sharply decreases as input size grows and finally slowly converges.
2 red-circled latency-optimal points given by our optimization both correspond to the turning point of the per-token latency curve, beyond which increasing the graph input size no longer improves hardware utilization and instead introduces additional padding overhead. Figure~\ref{fig:latency-all} further illustrates the  stair-step pattern relationship between latency and input size due to padding. 
It also shows that, under the latency-optimal configuration, selective recomputation (orange curve, $L_S=512$) achieves $2$--$3\times$ lower latency than ordinary prefill (blue curve, $L_P=128$) for processing non-prefix reusable tokens.

\paragraph{Mixed-Precision Quantization for KV Deviation Computation}
\label{mp-quant}
The KV deviation computation introduced in the selective KV recompute pipeline is precision-sensitive and input-dependent (dynamicity \textcircled{3} analyzed in Section~\ref{bg-reuse}). 
Thus, directly quantizing it to $\mathrm{INT8}$ $\mathrm{UINT16}$ like other intermediate activations can lead to significant accuracy degradation. 
To address this, we propose a mixed-precision selective KV recompute graph, where the deviation computation and Top-K comparison are all conducted on $\mathrm{FP16}$ precision. We further apply a scale factor $\frac{1}{\sqrt{D_{head}}}$ to scale down the intermediate deviation tensor to prevent $\mathrm{FP16}$ overflow. Such mixed-precision quantization successfully preserves the model precision, verified in the ablation study.

\subsection{Inter-Graph Design for Dynamic Reuse Pattern}
\label{static-util}
% The motivation of graph utilization optimization involves 2 major aspects: maximizing hardware utilization, such as NPU TCM capacity and matrix units (e.g., Qualcomm QNN HMX tiles), and minimizing padding overhead caused by irregular input shapes.
%%\paragraph{Dynamic programming chunk merge algorithm for efficient static graph utilization.}
%To address this inefficiency, we formulate a dynamic programming chunk merge algorithm to improve the static graph utilization. 

Within an input prompt, prefix reusable, non-prefix reusable, and new tokens may come interleavingly (dynamicity \textcircled{5}). 
They invoke distinct graph calls by default:
prefix reuse tokens incur no graph calls, non-prefix reusable tokens are processed by the selective recomputation graph, and new tokens are processed by the full prefill graph. A naive schedule which invokes static graphs solely according to token types may cause fragmentation between tokens of different types, often producing underfilled graph calls and wastes computation due to padding. To address this inefficiency under dynamic patterns, we propose a chunk merging algorithm based on dynamic programming to improve the static graph utilization. The key observation is that \textit{non-cached new tokens can be safely processed by the selective recomputation graph}, since they are always selected for computation. Our algorithm can dynamically merge new tokens into adjacent under-utilized selective recomputation graphs to improve efficiency.

Let $T[i]$ be the minimum latency to process tokens up to position $i$.
For the last chunk ending at $i$, we consider 2 choices: using the selective recomputation graph or using the prefill graph.
This yields the optimal substructure transition of dynamic programming:
  \begin{equation}
      T[i] = \min \bigl\{ T[i-l_s(i)] + t_s,\; T[i-l_p(i)] + t_p \bigr\},
    \label{eq:dp-chunk-merge}
  \end{equation}
where $t_s$ and $t_p$ are the per-call latency of the selective recomputation and prefill graphs, respectively.
Here, $l_s$ is the maximum suffix length ending at $i$ that can be absorbed by one selective recomputation graph call under graph capacity and recomputation-ratio constraint (i.e., at least $r$ non-prefix tokens can be recomputed). $l_p$ is the maximum suffix length ending at $i$ that can be processed by one prefill graph call under its capacity.
Intuitively, $l_s$ may include both non-prefix reuse tokens and a bounded number of adjacent new tokens, whereas $l_p$ simply packs as many trailing tokens as the prefill graph allows.
In Equation \ref{eq:dp-chunk-merge}, the first term $T[i-l_s(i)] + t_s$ is the optimum of the subproblem if the recomputation graph is used, while the second term $T[i-l_p(i)] + t_p$ corresponds to using the prefill graph.
The dynamic program therefore globally chooses whether each trailing chunk should be executed by recomputation or prefill, rather than making a purely local greedy decision.
The formal formulation of this constrained graph scheduling and complete pseudo code are provided in the supplementary material.

\begin{figure}[h]
    \centering
    \includegraphics[width=\linewidth]{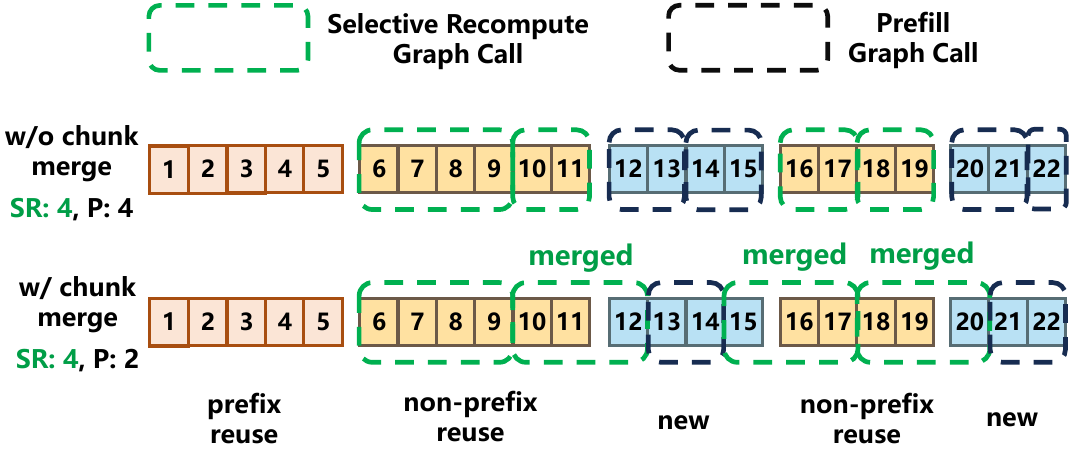}
    \caption{Chunk merging algorithm example, which saves 2 prefill graph calls and reduces $\approx 25\%$ latency.}
    \label{fig:chunk-merge}
\end{figure}

Figure~\ref{fig:chunk-merge} shows a toy example: when the recomputation graph processes up to 4 tokens, the prefill graph processes up to 2 tokens, and $r=50\%$, new tokens 12, 15, and 20 can be merged into adjacent recomputation calls. As a result, the number of prefill graph invocations is reduced from 4 to 2, yielding an approximate 25\% latency reduction by utilizing otherwise underfilled recomputation graphs.

\subsection{Generality of Our Design}
\label{sec:generality}

Our compute-side design is not specific to Qualcomm HTP.
It is applicable to mobile accelerators that expose ahead-of-time compiled static graphs over fixed-shape tensors, such as MediaTek APU and Apple ANE.
Under this execution model, the intra-graph mechanisms in \S\ref{static-dynamic}, including chunk--pad--mask and scatter--gather, can be expressed using standard tensor operators supported by these backends.
The inter-graph scheduler in \S\ref{static-util} depends only on the capacity and measured latency of the available pre-compiled graphs, and is therefore independent of a particular accelerator implementation.
Porting to a new backend requires recalibrating the graph sizes $L_P$ and $L_S$ and the hardware-specific alignment granularity following the procedure in \S\ref{static-dynamic}.
Thus, we expect migration to require backend integration and profiling rather than redesigning the compute-side mechanisms.

\section{On-Device Hierarchical KV Storage}
\label{storage-design}

\begin{figure}[t]
    \centering
    \includegraphics[width=0.85\linewidth]{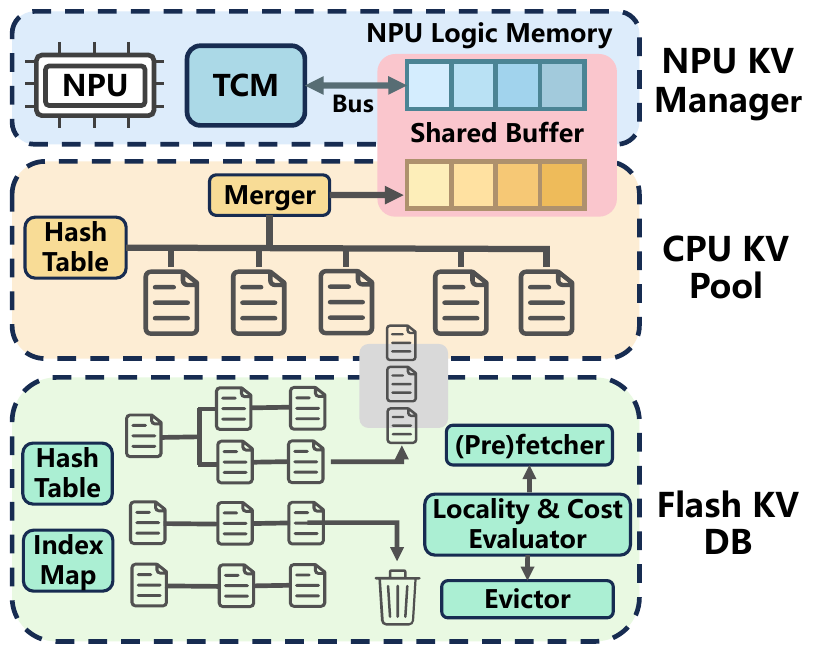}
    \caption{On-device hierarchical KV caching system.}
    \label{fig:kv_storage}
\end{figure}

To stably supply reusable KV tensors to NPU under the strict on-device memory bandwidth and capacity constraints, we design a hierarchical KV storage system upon the on-device memory hierarchy. 
Inside, the data structures in memory and flash storage leverage heterogeneous memory characteristics to balance reuse efficiency and data movement overhead.
Prefetch and eviction policies are also activated to mitigate cache misses across memory and storage tiers and consequently reduce end-to-end latency. The architectural design of KV caching system is illustrated in Figure \ref{fig:kv_storage}. 
The system majorly consists of 3 structures: NPU KV manager, CPU KV memory pool, and the flash KV DB.
During the workflow, KV tensors are efficiently matched and loaded from the flash DB and (pre)fetched to the CPU memory pool. When transmission signals of NPU are received, the tensors are further transferred to the CPU--NPU shared buffer for NPU to consume. 

\subsection{NPU-CPU In-Memory KV Manager}
\label{cpu-npu-mngr}

\paragraph{NPU KV Manager} 
A fixed-size (context-length) KV buffer on CPU--NPU shared buffer, typically on the order of $10^1$ MB is allocated by the NPU KV manager. 
This buffer contains the KV cache of the running session, where NPU directly loads KV tensors in this region with DMA into TCM and consumes it by the matrix unit. 
The manager controls its sharing behavior between CPU and NPU and updates the buffer throughout LLM executions.

\paragraph{CPU KV Memory Pool} 
A medium-sized KV pool is maintained in CPU memory, typically on the order of $10^2$ MB, and is implemented as a lightweight hash table based structure that indexes cached KV tensors using their corresponding underlying database row IDs. 
To save memory, the pool stores KV chunks that are either newly generated by the NPU or (pre)fetched from flash storage, rather than being a full tree-based structure as cloud-based existing works \cite{SGLANG, RAGCache}. 
Upon receiving a new input request, the system first queries the CPU KV pool before accessing flash storage to enable low-latency reuse. 
Eviction is triggered under system memory pressure, such as when available memory becomes limited due to contention from other mobile applications.

\subsection{Prefix and Non-Prefix Hybrid Structure in Flash}
\label{secondary}
We organize KV caches in on-device flash storage using a tree--hash--index hybrid structure implemented on top of SQLite~\cite{SQLite}, which jointly supports both prefix-based and non-prefix KV reuse while maintaining efficient lookup and moderate storage overhead. 
This storage layer is designed to operate at GB scale and can dynamically trigger eviction when requested by user or system under storage pressure.

Specifically, following prior work~\cite{SGLANG, RAGCache}, we construct a prefix tree to support efficient prefix reuse and hierarchical organization of cached KV entries. 
However, tree-based structures alone cannot efficiently support non-prefix reuse. 
To address this limitation, we augment the structure with both hash-based and semantic indexing mechanisms. For non-prefix reuse, each cached KV chunk is associated with a compact 64-bit hash code, enabling fast lookup of identical reusable chunks. 
For subchunk-level reuse, we additionally maintain lightweight semantic indices,
including sparse embedding based on extracted keywords and truncated dense embedding vectors. 
Dense embedding is only in RAG use cases where embedding is already computed and stored in the workflow. 
This structure enables fast localization of potential non-prefix chunk candidates that may contain reusable subchunk sequences, even when the text is not identical at chunk level. Linear-time fast longest common substring (LCS) algorithm is further applied to identify and finalize the reusable subchunks among the retrieved candidates.

%A linear-time approximate LCS (Longest Common Substring) is performed to finalize the reuse chunks among the candidates.

Together, the prefix tree for prefix reuse, and hash table and semantic index for non-prefix reuse, form a unified flash storage organization that supports both types of KV reuse across diverse long-context workloads. 
The detailed database schema and storage organization are described in the supplementary material.

\subsection{Prefetch and Eviction} \label{eviction}
%Prefetch and eviction policies are critical for reducing cache misses and end-to-end latency. To improve cache efficiency across memory and storage tiers, we design asynchronous prefetch and cost-aware eviction mechanisms.

\paragraph{Asynchronous Prefetch}
To improve prefix hit rate in the CPU memory pool, the system performs asynchronous prefetch between flash and main memory. During idle I/O periods, prefix-reusable chunks with high predicted reuse likelihood are proactively loaded into memory to reduce future access latency. To avoid unnecessary memory pollution, the prefetch policy selectively prioritizes chunks that are likely to participate in future prefix reuse or are strongly correlated with recently accessed chunks.

\paragraph{Cost-Aware Eviction}
We jointly consider chunk reusability and eviction overhead to compute an overall eviction score for each chunk. Chunks that are both highly reusable and expensive to recompute or reload are preferentially retained under constrained memory and storage budgets, while lower-value chunks are evicted when necessary. This improves cache utilization and reduces end-to-end TTFT.

For chunks in the CPU memory pool, eviction overhead mainly corresponds to reload latency from flash storage for prefix-reusable chunks, while the cost for non-prefix-reusable chunks is negligible because their loading latency can largely overlap with execution (\S~\ref{store-npu}).

For chunks stored in flash, eviction may incur not only the recomputation cost of the evicted chunk itself, but also additional overhead from disrupting the prefix reuse chain. 
In particular, evicting a chunk on a prefix path can cause its descendant chunks to lose prefix reusability (Figure~\ref{fig:eviction-cost}), thereby increasing future recomputation cost. 
Our eviction policy explicitly models this cascading effect when making retention decisions.

\begin{figure}[!t]
    \centering
    \includegraphics[width=0.63\linewidth]{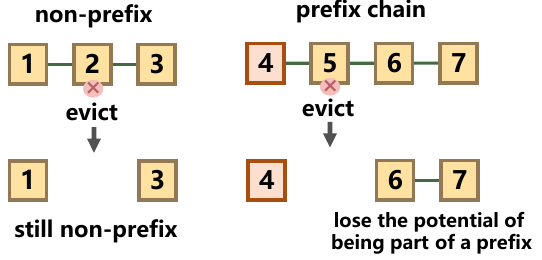}
    \caption{Evicting chunks on a prefix chain breaks future prefix reuse for subsequent chunks.}
    \label{fig:eviction-cost}
\end{figure}

\section{Compute--Storage Pipeline Overlap}
\label{pp}
Due to the limited bandwidth of on-device memory hierarchies, optimizing compute and storage independently is insufficient. 
Our system overlaps KV loading, rerotation, storing, and NPU execution through asynchronous pipeline parallelism to hide data-movement latency.

\subsection{KV Loading Latency Hiding}
\label{store-npu}

During KV preparation, KV tensors are retrieved from storage and rerotated for non-prefix reuse before being consumed by the NPU.

\begin{figure}[!h]
    \centering
    \subfigure[CPU KV rerotation.]{
    \includegraphics[width=0.42\linewidth]{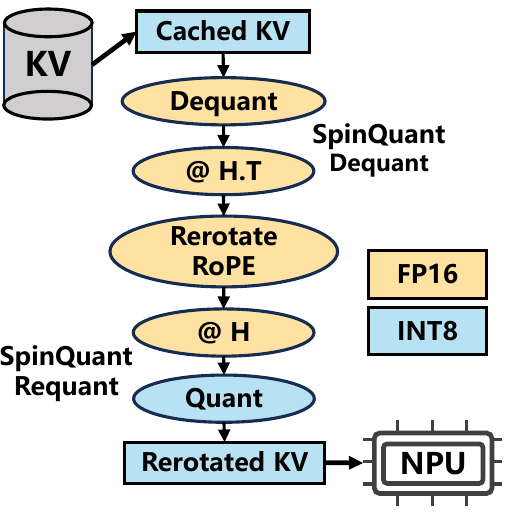}
    \label{fig:CPU-rerot}}
    \hfill
    \subfigure[IO--CPU--NPU pipeline parallelism of KV preparation and LLM inference.]{
    \includegraphics[width=0.526\linewidth]{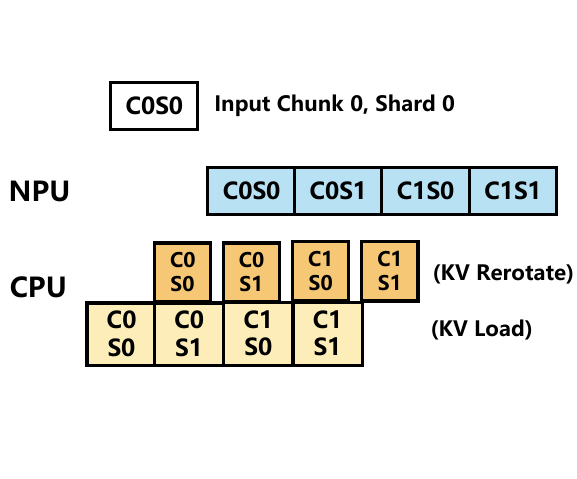}
    \label{fig:CPU-NPU-PP}}
    \caption{IO--CPU--NPU overlapping pipeline.}
    \label{fig:KV}
\end{figure}

\paragraph{IO-CPU-NPU Overlapping}
To enable non-prefix KV reuse, the rotary embedding (RoPE) of K tensors is rerotated according to their original positions in cached prompts and their new positions in incoming queries~\cite{CacheBlend,ChunkAttention,PromptCache}. 
In the quantized on-device workflow, this process additionally involves KV dequantization and requantization with Hadamard transforms introduced by 8-bit SpinQuant~\cite{SpinQuant}. 
Since rerotation is memory-intensive\footnote{Fast Walsh--Hadamard transformation has $O(n\log n)$ complexity and is memory-intensive in our setting.}, it is assigned to the CPU and overlapped with compute-intensive NPU execution. 
As illustrated in Figure~\ref{fig:CPU-rerot}, both I/O-intensive KV loading and memory-intensive rerotation are executed asynchronously on the CPU to avoid stalling the NPU.

\begin{figure}[!h]
    \centering
    \subfigure[Shard-level parallelism reduces prefix loading time 4$\times$.]{\includegraphics[width=0.44\linewidth]{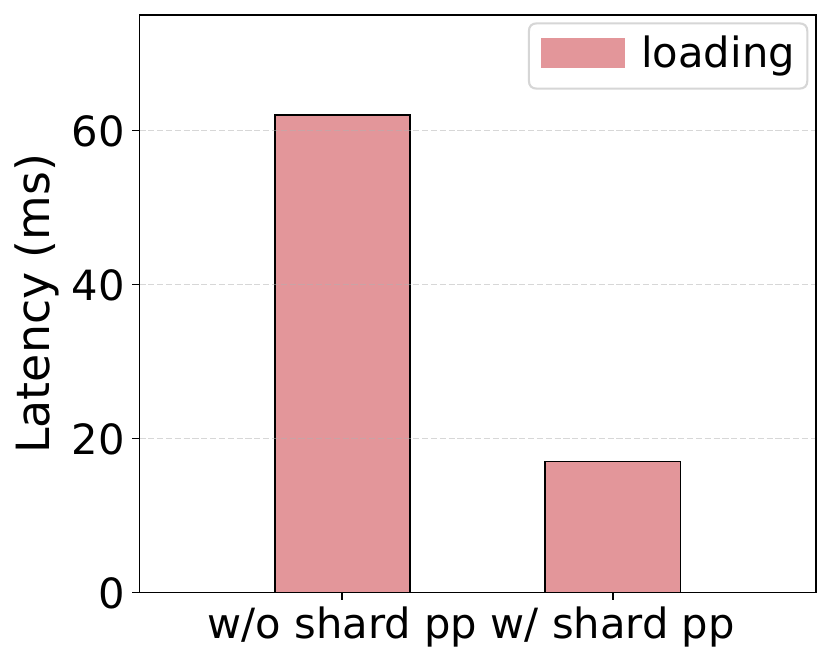}}
    \hfill
    \subfigure[IO--CPU-NPU parallelism makes data preparation latency invisible to NPU execution.]{\includegraphics[width=0.52\linewidth]{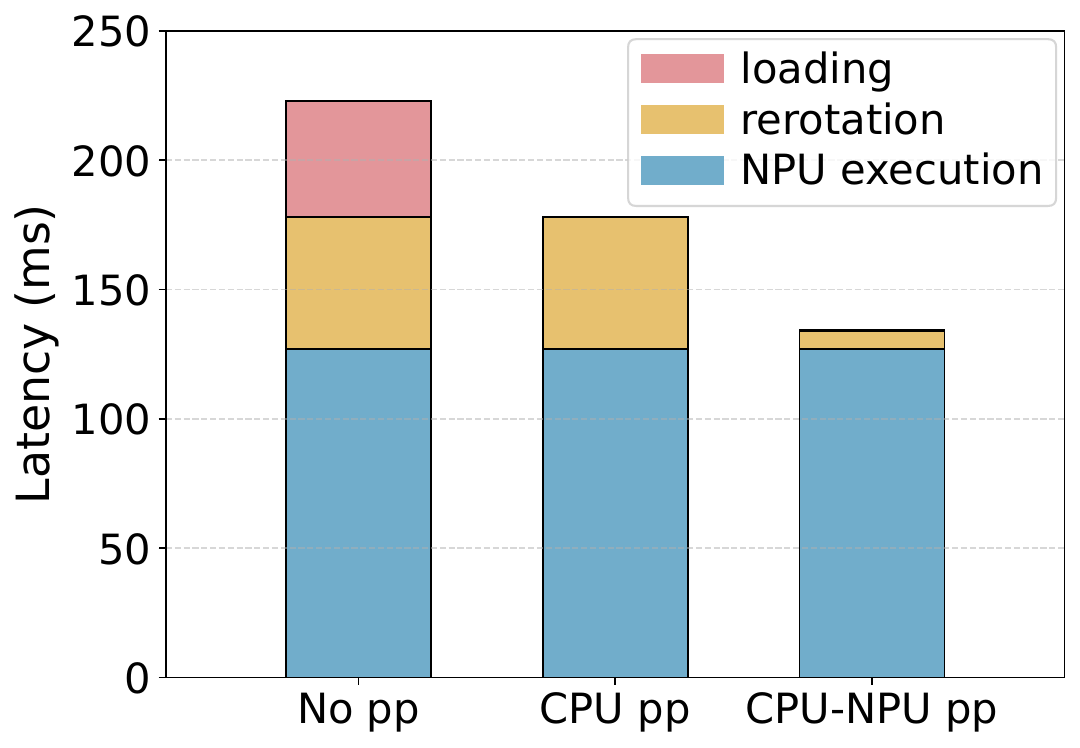}}
    \caption{Hiding KV loading and rerotation latency via pipeline parallelism. (Qwen3-1.7B on Xiaomi 15 Pro)}
    \label{fig:overlap-load}
\end{figure}

\paragraph{Chunk-Shard Two-Dimensional Pipeline Parallelism}

Mobile NPUs execute LLMs in multiple model shards while prompts are processed in chunks~\cite{mllm}, enabling pipeline parallelism along both dimensions. 
As illustrated in Figure~\ref{fig:CPU-NPU-PP}, while the NPU executes shards of Chunk~0, the CPU concurrently prepares KV tensors for upcoming shards of Chunk~0 and subsequent Chunk~1. 
Figure~\ref{fig:overlap-load} shows that incorporating shard-level parallelism reduces the non-overlapped loading latency of prefix chunks and the first non-prefix chunk by 4$\times$. 
Overall, this two-dimensional pipeline substantially increases overlap opportunities and effectively hides KV loading and rerotation latency.

\subsection{KV Storing Latency Hiding}

Synchronously serializing KV tensors from the CPU memory pool to flash storage can introduce substantial user-visible latency due to limited secondary-storage bandwidth. 
To mitigate this overhead, we adopt a lazy and asynchronous KV persistence mechanism.

Instead of synchronously writing KV tensors to flash immediately after generation, newly produced KV chunks are first retained in the CPU memory pool and marked as dirty entries, while the actual serialization and storage operations are deferred and executed asynchronously by background threads when the I/O bus is idle.
For example, KV storing can be overlapped with decoding. 
In addition, write operations are batched whenever possible to reduce flash I/O amplification and metadata overhead.
This design effectively hides most flash write latency from the critical inference path.

%\section{Implementation}

\section{Evaluation}
\label{evaluation}

\subsection{Experimental Setup}\label{setup}

\paragraph{Use Cases and Datasets}
We take the following three representative on-device LLM use cases.

%: document QA, long-context history QA, and agent skill use.

\begin{itemize}
    \item 
For \emph{document QA}, we randomly sample 400 easy-level queries from HotpotQA~\cite{HotpotQA}.
For each query, we split its associated context into 1024-character chunks to build the context chunks document database, and retrieve the top-6 chunks based on the L2 distance between embeddings generated by Qwen3-Embedding-0.6B~\cite{Qwen3-Embedding}.

\item For \emph{long chat history QA}, we use LoCoMo-MC10~\cite{LoCoMo‑MC10}, derived from LoCoMo~\cite{LoCoMo}.
Two sessions of very long chat history with 74 single-hop QA questions are selected to evaluate reuse under long conversation contexts. Top-6 related dialog chunks are retrieved and combined into the prompt, with the same setting as HotpotQA.

\item For \emph{agent skill use}, we construct a skill-oriented evaluation set from SkillsBench~\cite{SkillsBench}, including 10 shared skills and 39 associated task instructions.
Each testing sample contains a reusable skill description and a task instruction, enabling evaluation under shared skill contexts.
\end{itemize}

These workloads cover both contiguous prefix reuse and more general non-prefix reuse patterns, where reusable content may appear as retrieved document chunks, historical conversation segments, or shared skill descriptions. 

\paragraph{Devices}  Our evaluation includes one mid-range device, Meizu 21, and two high-end devices, Xiaomi 15 Pro and Honor Magic 8. Their hardware specifications are summarized in Table~\ref{tab:devices}. All experiments are conducted using the Qualcomm Hexagon NPUs available on these devices.

%Test devices involve 1 mid-range device: Meizu 21 and 2 high-end devices: Xiaomi 15 Pro and Honor Magic 8. The hardware specification is illustrated in Table \ref{tab:devices}. All our experiments are conducted on their Qualcomm Hexagon NPUs.

\begin{table}[h]
    \centering
    \caption[Mobile devices used in our experiments.]{Mobile devices evaluated in our experiments.}
\resizebox{\linewidth}{!}{
    \begin{tabular}{cccccc}
    \toprule[1.5pt]
        {\bf Device}  & {\bf SoC} & {\bf Mem} & {\bf CPU} & {\bf GPU} & {\bf NPU} \\
         \midrule
         \multirow{2}{*}{Meizu 21} & Snapdragon  & 12GB & 1*Cortex-X4+5*A720   & Adreno & Hexagon \\
          & 8 Gen 3 & +256GB  &  +2*A520 & 750  &  V75  \\
         \midrule
         \multirow{2}{*}{Xiaomi 15 Pro} & Snapdragon & 16GB  &  \multirow{2}{*}{8*Oryon}  & Adreno & Hexagon  \\
         & 8 Elite & +512GB &  & 830  & V79    \\
         \midrule
         \multirow{2}{*}{Honor Magic 8} & Snapdragon  & 12GB & \multirow{2}{*}{8*Oryon}   & Adreno  & Hexagon  \\
          & 8 Elite Gen 5 & +512GB  &   & 840 &  V81  \\
    \bottomrule[1.5pt]
    \end{tabular}
    }
    \vspace{0.5em}
    \label{tab:devices}
    \vspace{-1em}
\end{table}

%\footnotetext{All the Android CPU start with A or X stands for ARM Cortex-A or ARM Cortex-X. Some of the iPhone information is inaccessible and left blank.}

\begin{figure*}[ht]
    \centering
    \subfigure[HotpotQA: our design reduces 32--56\% TTFT.]{
        \includegraphics[width=0.31\linewidth]{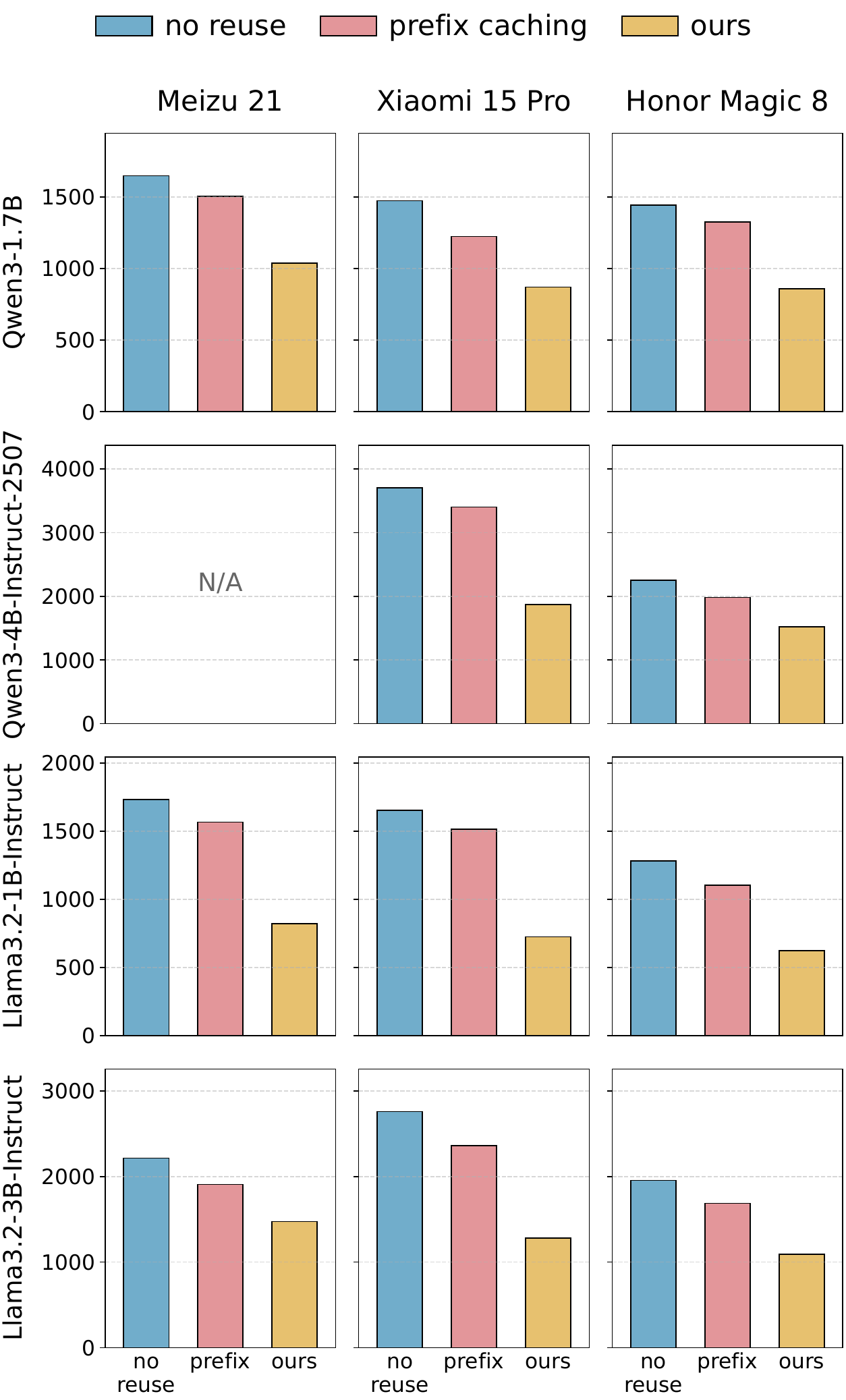}
    }
    \hfill
    \subfigure[LoCoMo: our design reduces 42--65\% TTFT.]{
        \includegraphics[width=0.31\linewidth]{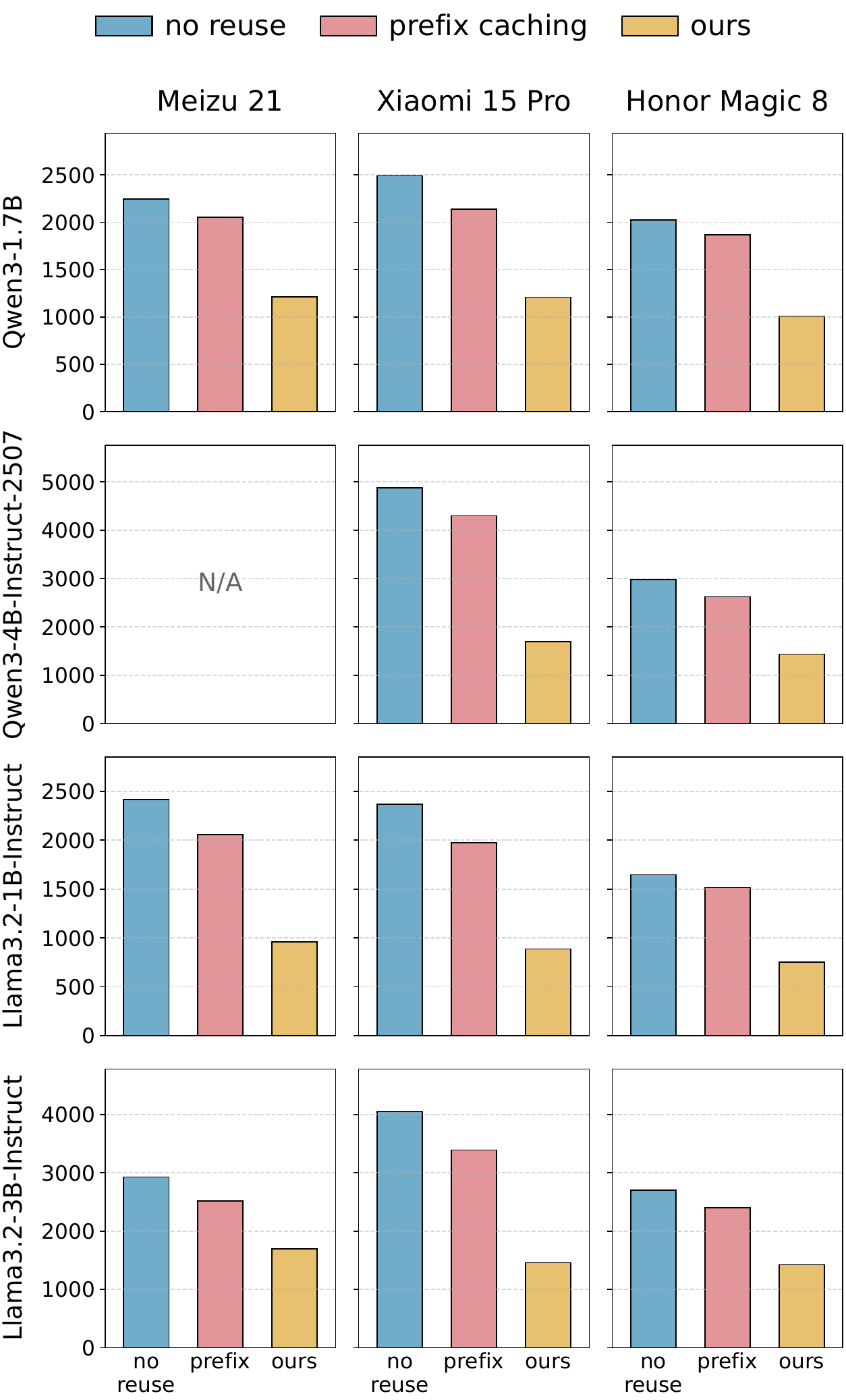}
    }
    \hfill
    \subfigure[SkillsBench: ours design reduces 68--81\% TTFT.]{
        \includegraphics[width=0.31\linewidth]{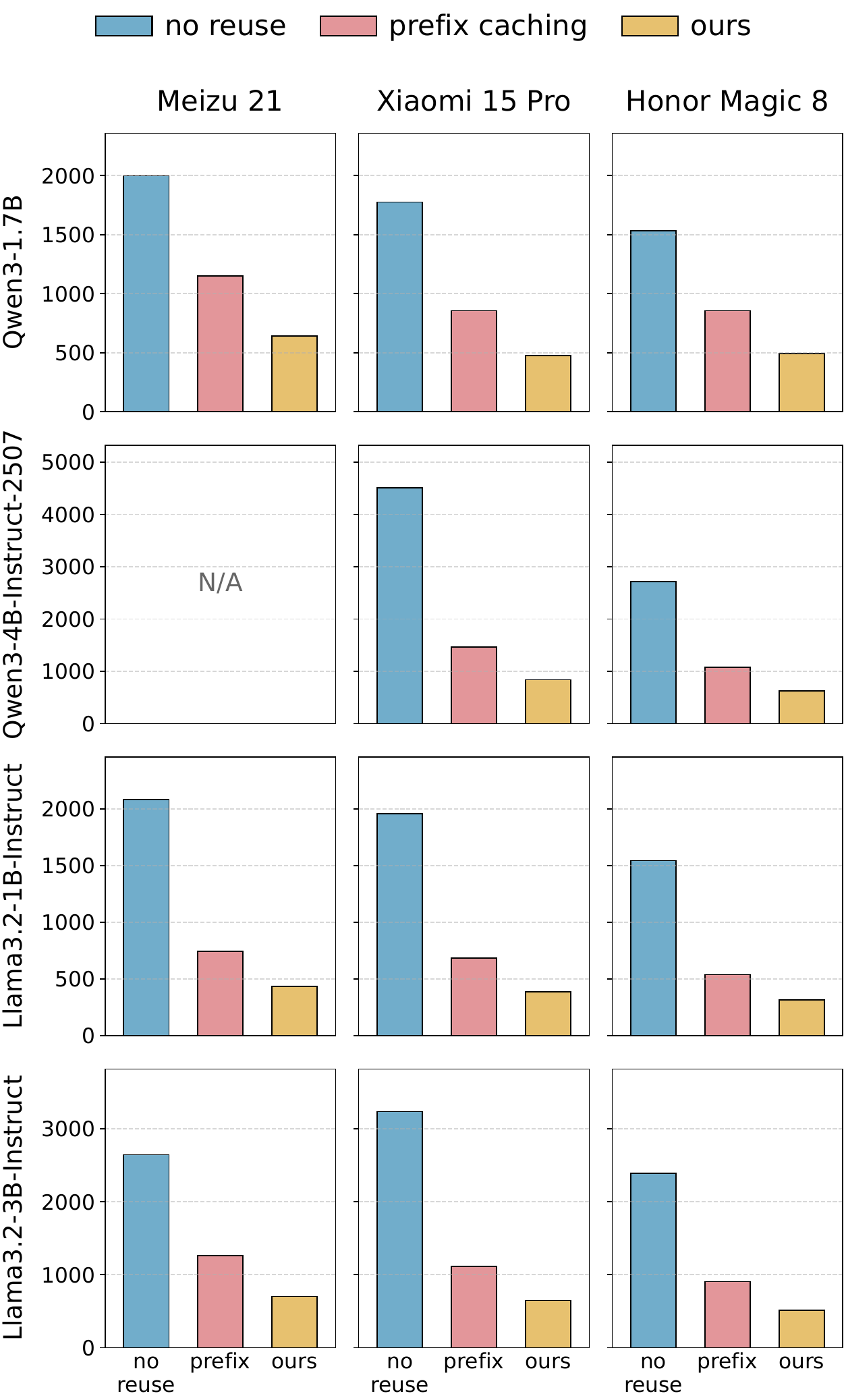}
    }
    \caption{End-to-end TTFT across 3 workloads, 4 LLMs, and 3 smartphones. Our system consistently reduces TTFT compared with no reuse (original ExecuTorch) and prefix caching. The TTFT reduction ranges in each subcaption are measured against original ExecuTorch across all available model-device pairs.}
    \label{fig:e2e-grid}
\end{figure*}

\paragraph{Models} We take Qwen3-1.7B and Qwen3-4B \cite{Qwen3}, and Llama3.2-1B-Instruct and Llama3.2-3B-Instruct \cite{Llama3, Llama3.2}. Models are majorly quantized into Int4.

\paragraph{Base Engine and NPU Backend}
We adopt ExecuTorch~\cite{ExecuTorch} as the base mobile LLM engine, which provides state-of-the-art performance on NPUs together with a stable and extensible deployment framework for on-device LLM inference. We target Qualcomm HTP NPUs, motivated by their broad deployment and mature public software stack.

%together with a stable and extensible deployment framework for on-device LLM inference. 
%The current implementation targets Qualcomm HTP NPUs due to their large market share and mature public software stack. 
%Support for additional mobile inference engines and NPU platforms will be explored in future work.

%%\emph{no reuse}, \emph{prefix caching}, and \emph{full reuse}. The baseline configurations are summarized as follows:

\paragraph{Baselines}
We compare against three baselines representing different KV reuse strategies: 
\begin{itemize}
    \item \emph{no reuse}: the original ExecuTorch ~\cite{ExecuTorch} implementation without KV reuse.

    \item \emph{prefix caching}: an implementation with tree-based prefix caching, reproducing SGLANG Radix-Tree prefix reuse on device~\cite{SGLANG}.

    \item \emph{full reuse}: an implementation that directly reuses both prefix and non-prefix KV chunks without selective recomputation, reproducing Prompt Cache full reuse on device~\cite{PromptCache}.
\end{itemize}

The KV storage backend used by the \emph{prefix caching} and \emph{full reuse} baselines is also implemented by us on top of SQLite to ensure a fair comparison under the same on-device storage stack. 
We do not directly compare against cloud-based systems, such as vLLM, SGLang, LMCache, or RAGCache, because they are designed for GPU server environments and do not support mobile hardware.

%The KV storage backend used by the \emph{prefix caching} and \emph{full reuse} baselines is also implemented by us on top of SQLite. We do not directly compare against vLLM, SGLang, LMCache, or RAGCache because they do not support mobile devices runtime.

\paragraph{Metrics}
For all three use cases, we report the average TTFT (ms) over representative datasets to evaluate system efficiency. 
In addition, to validate the effectiveness of our CacheBlend style selective recomputation algorithm for non-prefix reuse on mobile devices, 
we evaluate generation quality on the document QA and chat-history QA workloads and report the corresponding accuracy. 
We also report the average system power and total energy consumption during prefill, measured via Android BatteryManager, to evaluate whether latency reductions translate into energy savings.

\paragraph{Key Configurations}
The graph sizes of the NPU selective recomputation graph and prefill graph are set to the latency-optimal values identified in \S~\ref{static-dynamic}: $L_S=512$ and $L_P=128$. 
To preserve generation quality while fully utilizing the HMX GEMM tiles~\cite{llm-npu}, the recomputation ratio is set to $r=0.25$, ensuring that intermediate tensor shapes remain multiples of 32.

\subsection{End-to-End Improvements}
\label{sec:e2e}

\begin{figure}[h]
    \centering
    \includegraphics[width=0.9\linewidth]{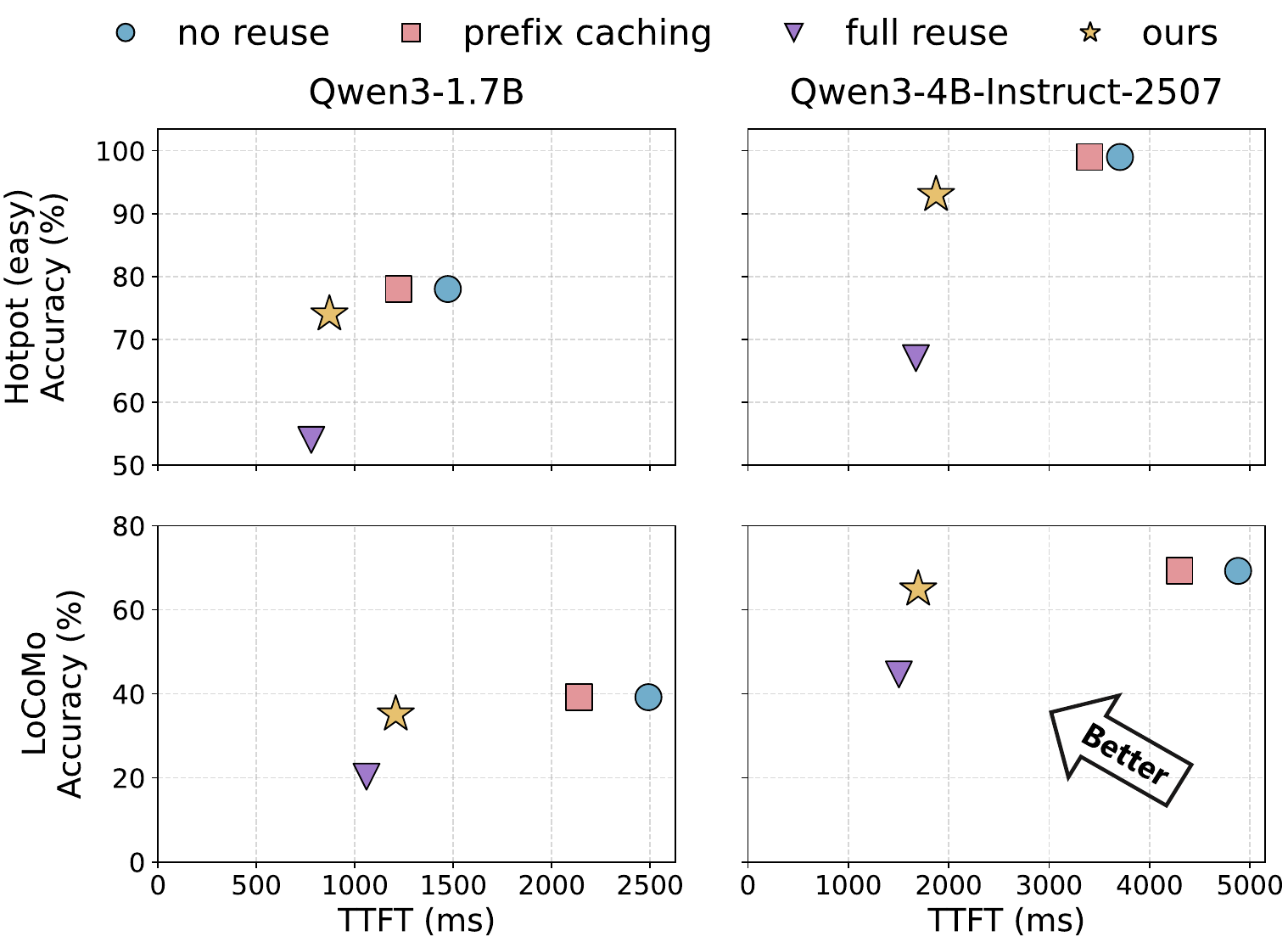}
    \caption{Quality--latency tradeoff on HotpotQA and LoCoMo using Qwen3-1.7B and Qwen3-4B on Xiaomi 15 Pro. Our system (yellow star) reduces TTFT substantially while preserving generation quality.}
    \vspace{-0.5em}
    \label{fig:ACC-TTFT}
\end{figure}

\paragraph{Reduced TTFT with Negligible Quality Degradation}
Figure~\ref{fig:ACC-TTFT} illustrates the quality--latency tradeoff on HotpotQA and LoCoMo using Qwen3-1.7B and Qwen3-4B on Xiaomi 15 Pro. 
The ideal operating region lies in the upper-left corner, corresponding to lower TTFT and higher task accuracy. 

Compared with \emph{no reuse} and \emph{prefix caching}, our system substantially shifts the operating point toward this region, reducing TTFT by approximately 40\%--60\% with only $\leq 4\%$ accuracy degradation. 
Although \emph{prefix caching} preserves full precision, its latency improvement is limited because it only exploits reusable contiguous prefixes. 

In contrast, the more aggressive \emph{full reuse} achieves slightly lower TTFT but incurs significant quality degradation (approximately 30\%) due to direct non-prefix KV reuse without selective recomputation. 
Moreover, \emph{full reuse} is only 12\% faster than our approach because its rerotation overhead cannot be effectively hidden through pipeline overlap and becomes exposed on the critical path. 
Therefore, we exclude \emph{full reuse} from the subsequent evaluations.

\paragraph{Efficiency Evaluation across Tasks, Models, and Devices}
Figure~\ref{fig:e2e-grid} compares TTFT among \emph{no reuse}, \emph{prefix caching}, and \emph{ours} across different workloads, LLMs, and smartphones. 
Across all configurations, our system consistently achieves the lowest TTFT, delivering 1.5$\times$--5.4$\times$ prefill speedup over \emph{no reuse} and 1.3$\times$--2.5$\times$ speedup over \emph{prefix caching}.

In the agent skill-use workload, reuse patterns are largely prefix-dominated because multiple tasks share the same skill descriptions and tool instructions. 
As a result, both \emph{prefix caching} and our system reduce TTFT by more than 50\%. 
Our system further improves over \emph{prefix caching} through optimized prefetching and compute--storage pipeline overlap.

In contrast, document QA and long-chat workloads exhibit substantially more non-prefix reuse. 
Retrieved document chunks or dialog histories are dynamically assembled into prompts with different orders and combinations, making reusable prefixes cache hits much more difficult. 
Under such patterns, \emph{prefix caching} achieves only limited gains (1.1$\times$-1.2$\times$), especially on mobile devices where constrained memory capacity causes frequent Radix-Tree eviction and prevents caching all prompt combinations. 
In contrast, by supporting both prefix and non-prefix KV reuse through selective recomputation, our system continues to achieve substantial TTFT reduction, providing 1.5$\times$--2.9$\times$ speedup in these workloads.

\paragraph{Energy Reduction}
We measure prefill energy consumption on Xiaomi 15 Pro with Qwen3-4B-Instruct-2507 across the three datasets and compare our
system against \emph{no reuse}.
Our system maintains a comparable average power draw to \emph{no reuse}: 6.34~W vs.\ 6.68~W, indicating that the additional KV management operations do not materially increase system power.
Consequently, the reduction in prefill latency directly translates into lower energy consumption.
As shown in Table~\ref{tab:energy}, our system reduces prefill energy by 52\% on HotpotQA, 67\% on LoCoMo, and 77\% on SkillsBench. 
These reductions closely track the corresponding TTFT improvements in \S7.2, confirming that our system substantially reduces both latency and energy consumption.

\begin{table}[h]
\centering
\caption{Prefill energy reduction measured on Xiaomi 15 Pro with Qwen3-4B-Instruct-2507 across 3 datasets.}
\label{tab:energy}
\resizebox{0.75\linewidth}{!}{\begin{tabular}{lccc}
\toprule
\textbf{Energy (J)} & \textbf{HotpotQA} & \textbf{LoCoMo} & \textbf{SkillsBench} \\
\midrule
No Reuse & 24.76 & 32.61 & 18.02 \\
\textbf{Ours} & 11.89 & 10.75 & 4.13 \\
\midrule
 & -52.0\% & -67.0\% & -77.1\% \\
\bottomrule
\end{tabular}
}
\end{table}

\subsection{Sensitivity Analysis}

For a better understanding of on-device KV reuse mechanism, we analyze how the reuse ratio and prompt length affect the end-to-end prefill latency. We report the latency reduction range across all available model-device pairs.

\begin{figure}[h]
    \centering
    \includegraphics[width=\linewidth]{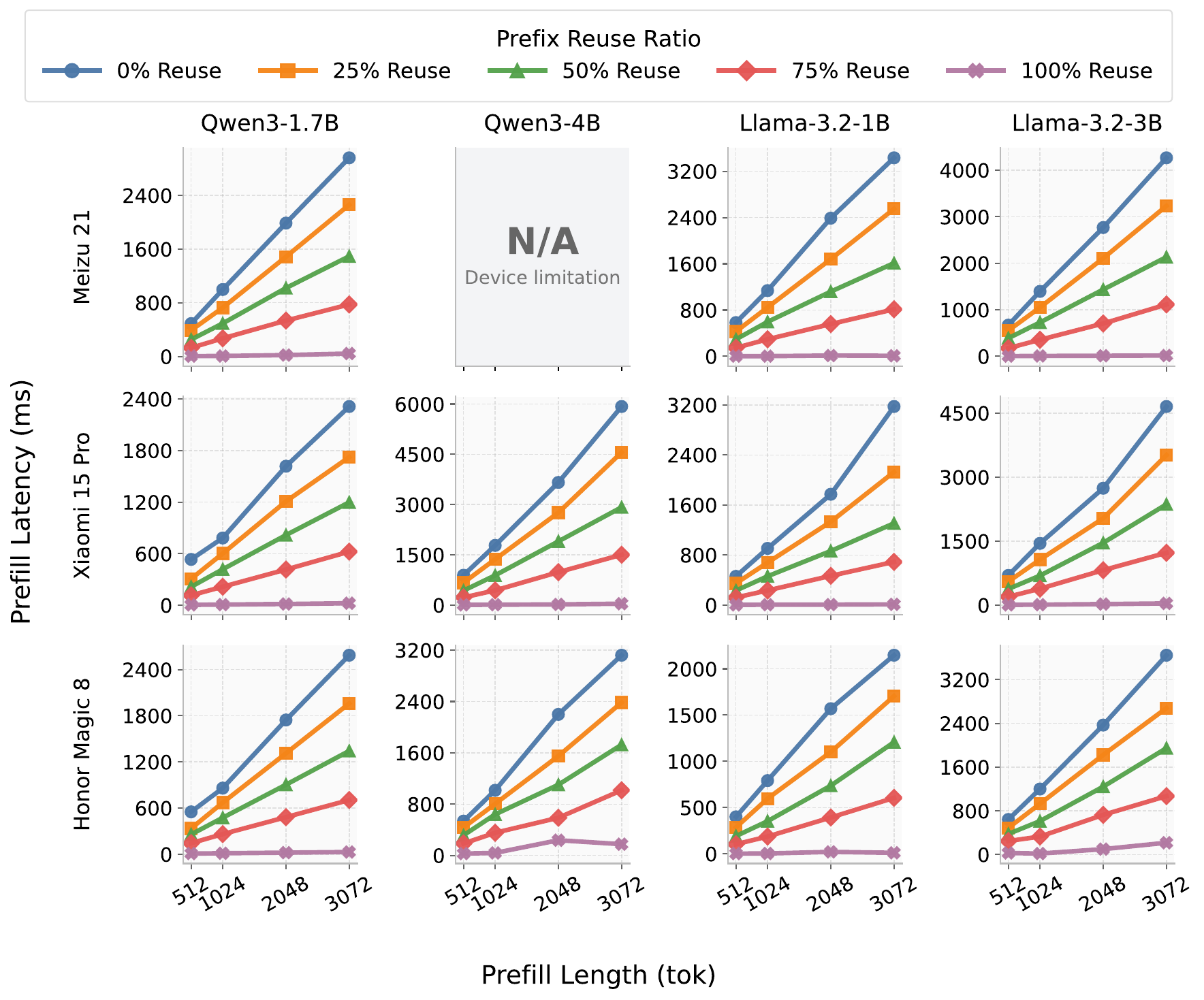}
    \caption{TTFT under different prefix reuse ratios and input lengths across 3 mobile devices and 4 models.}
    \vspace{-0.5em}
    \label{fig:sensitivity-prefix}
\end{figure}

\paragraph{Varying Prefix Reuse Ratio.}
Figure~\ref{fig:sensitivity-prefix} shows that prefix reuse consistently reduces prefill latency as the reuse ratio increases, and the benefit becomes more significant for longer prompts. At 3072 tokens, the full-prefill baseline takes 2.15--5.93~s across different models and devices. With 75\% prefix reuse, our system reduces the prefill latency by 1.54--4.43~s, corresponding to a 67.4--78.3\% reduction. When the reuse ratio reaches 100\%, the latency reduction improves to 94.1--99.7\%, leaving 9--215~ms residual prefill latency. 
These results indicate that when reusable KV cache forms a contiguous prefix, our system can reuse with little extra overhead, while the speedup is more obvious for higher hit ratio.

\begin{figure}[h]
    \centering
    \includegraphics[width=\linewidth]{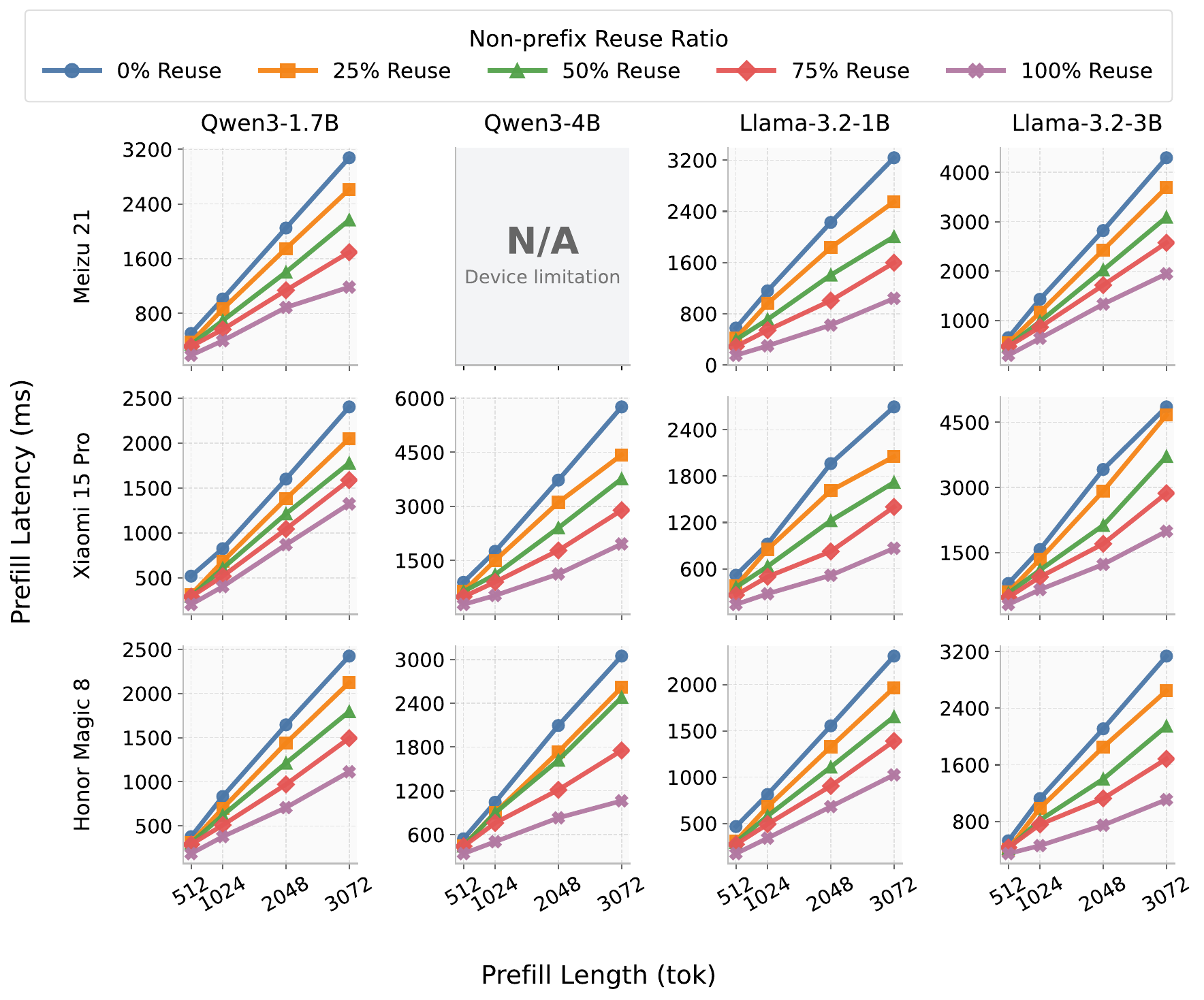}
    \caption{TTFT under different non-prefix reuse ratios and input lengths across 3 mobile devices and 4 models.}
    \vspace{-0.5em}
    \label{fig:sensitivity-nonprefix}
\end{figure}

\paragraph{Varying Non-Prefix Reuse Ratio}
Figure~\ref{fig:sensitivity-nonprefix} shows that non-prefix reuse also reduces end-to-end prefill latency, although the residual latency is higher than prefix reuse. 
At 3072 tokens, 100\% non-prefix reuse reduces the prefill latency by 1.08--3.80~s, corresponding to a 44.9--67.9\% reduction over the no-reuse baseline. 
The remaining latency, 0.87--1.99~s, is due to non-prefix reuse and selective recompute overhead. 
Nevertheless, the end-to-end prefill latency remains substantially lower than full prefill, showing that the saved model computation outweighs the extra reuse-management overhead.

\begin{figure*}[t]
    \centering
    \includegraphics[width=\linewidth]{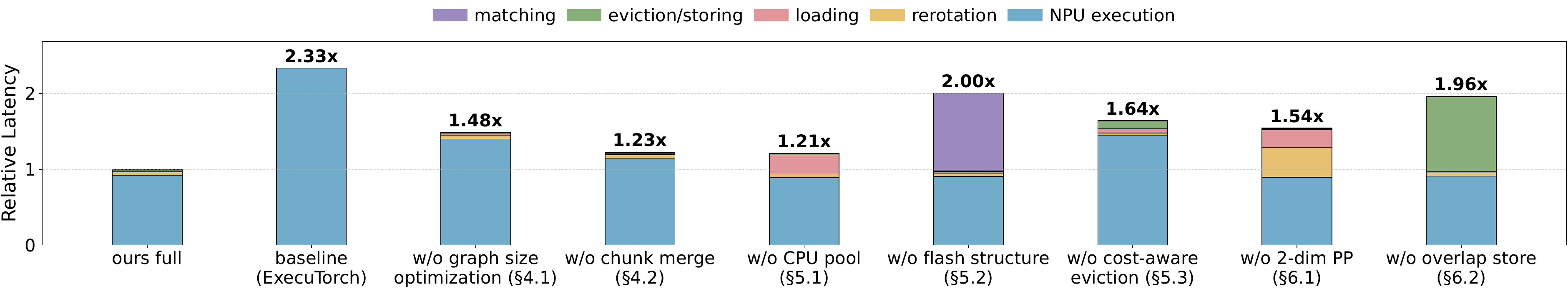}
    \caption{Ablation study and latency decomposition on Qwen3-1.7B, LoCoMo dataset, Xiaomi 15 Pro. Disabling individual components from our design increases TTFT by 1.21--2.33$\times$.}
    \vspace{-0.5em}
    \label{fig:ablation}
\end{figure*}

Overall, the results demonstrate that our method is consistently effective across different mobile devices and LLMs. Higher reuse ratios lead to larger latency reductions, and longer prompts further amplify the benefit of KV reuse. Prefix reuse provides the largest improvement due to its contiguous-cache structure, while non-prefix reuse offers a more general reuse capability with moderate additional overhead.

\subsection{Ablation Study}
\label{albation}
We decompose end-to-end latency into five components: prefix/non-prefix matching, eviction and storage, KV loading, KV rerotation, and NPU execution. 
Under the full system configuration, NPU execution dominates TTFT, accounting for more than 95\% of total latency, indicating that the overhead introduced by KV reuse is largely hidden and amortized.

As is illustrated in Figure \ref{fig:ablation}, removing individual system designs introduces noticeable overheads at different stages. In particular, disabling graph-size optimization and the chunk-merge algorithm reduces graph utilization efficiency and increases overall latency by 48\% and 23\%, respectively. 
Removing the CPU memory pool increases KV loading latency from flash storage, adding approximately 20\% latency. 
Replacing our hybrid structure in flash storage with a simple SQLite BLOB-based DB with naive structure increases matching overhead by an order of magnitude and nearly doubles overall latency. 
Disabling the eviction policy reduces the hit rate of non-prefix-reusable chunks by approximately threefold, resulting in a 60\% latency increase. 
Without IO--CPU--NPU pipeline parallelism, KV loading and rerotation can no longer overlap with NPU execution, increasing latency to 1.5$\times$. 
Similarly, disabling asynchronous KV storage exposes synchronous flash serialization overhead on the critical path and nearly doubles latency.

Overall, these results demonstrate that each major component contributes substantially to reducing end-to-end latency and improving KV reuse efficiency.

\section{Related Works}

\paragraph{Cloud-Based KV Reuse and Caching Systems} 
Cloud-based KV caching systems focus on organizing reusable KV caches across requests and cloud-based storage hierarchies. 
SGLANG~\cite{SGLANG} and RAGCache~\cite{RAGCache} employ prefix-aware tree structures, namely Radix Tree and knowledge tree, to store reusable prefixes. 
For non-prefix reuse, CacheBlend~\cite{CacheBlend} introduces a distributed KV storage and retrieval mechanism. 
More generally, LMCache~\cite{LMCache} provides a unified KV-cache layer for cache lookup, migration, and cross-engine sharing. 
Strata~\cite{Strata} studies hierarchical context caching for long-context serving, while Mooncake~\cite{Mooncake} advocates a KV-cache centric architecture that trades additional local storage for reduced recomputation. 
InfiniGen~\cite{InfiniGen} further accelerates KV access through essential-KV-only prefetching. 

\paragraph{On-Device LLM Acceleration} 
Existing approaches for accelerating on-device LLM inference primarily focus on optimized GEMM kernels, heterogeneous computing, quantized execution, and model sparsity.

ExecuTorch~\cite{ExecuTorch} provides high-performance NPU backends for Apple, MediaTek, and Qualcomm platforms through extensible and stable interfaces. 
MNN~\cite{MNN-LLM,Walle} improves mobile CPU and GPU inference through optimized KV-cache and weight layouts that enhance memory locality. 
MediaPipe~\cite{MediaPipe} (built on LiteRT~\cite{LiteRT}) and MLC-LLM~\cite{MLC-LLM} (built on TVM~\cite{TVM}) primarily optimize GPU-based prefill, while mllm~\cite{mllm} accelerates with optimized NPU GEMM kernels. 
SmartMem~\cite{SmartMem} reduces memory overhead by eliminating unnecessary NC4HW4 GPU layout transformations self-attention and GEMM execution. 
NPU flash attention and mixed-precision GEMM optimization~\cite{llm-npu} are also explored in recent works.
To better utilize memory bandwidth,  HeteroLLM~\cite{HeteroLLM} introductions GPU--NPU co-execution parallelism.  
T-MAC~\cite{T-MAC} and T-MAN~\cite{T-MAN} leverage table-lookup based quantized kernels for efficient CPU and NPU inference. 
PowerInfer 2~\cite{PowerInfer-2},  Apple Intelligence~\cite{LLM-in-a-Flash}, and Neuralink~\cite{Neuralink} exploit activation sparsity for efficient weight offloading.

Despite these extensive optimization efforts, KV reuse remains unexplored in prior on-device LLM systems.

\paragraph{On-Device Tensor Storage Systems} 

SQLite~\cite{SQLite}, a widely used on-device database system, provides BLOB support for tensor storage but lacks tensor-specific optimizations. 
Existing systems such as MicroNN~\cite{MicroNN} and MobileRAG~\cite{MobileRAG} focus on mobile vector databases, prioritizing embedding storage and building fast indices for vector approximate nearest neighbor (ANN) searches. 
Walle~\cite{Walle} and SFSL~\cite{SFSL} provide updateable FlatBuffers-based~\cite{FlatBuffers} tensor-storage systems for fixed-sized model weights and embeddings, but inefficient for dynamic-sized KV tensors.
For KV storage, SparKV~\cite{SparKV} studies device--cloud KV transfer for mobile LLM inference, while MobiLoRA~\cite{MobiLoRA} introduces an on-device KV-cache design for multi-LoRA scenarios, but both consider intra-session runtime small-scale KV storage only.

However, persistent and dynamically updateable hierarchical KV storage architectures tailored for on-device KV reuse have not been studied in existing work.

%However, persistent and updateable on-device hierarchical KV storage structures specifically designed for KV reuse remain largely unexplored.

\section{Conclusion}
In this work, we have designed and built an on-device KV reuse system to accelerate LLM inference on mobile NPUs. At the compute layer, we have transformed the dynamic selective recomputation workflow of non-prefix reuse into efficient static graphs for mobile NPUs. At the storage layer, we have designed a hierarchical KV caching system with hybrid indexing structures for efficient matching and reuse of prefix and non-prefix KV tensors. Compute and storage are further pipelined and overlapped to hide system overheads. Our implementation on Qualcomm Hexagon NPUs demonstrates that efficient non-prefix KV reuse is practical on commodity smartphones. Across representative workloads, LLMs, and devices, our system consistently reduces TTFT over no reuse and prefix-only caching baselines, while maintaining negligible quality degradation. More broadly, this work provides a foundation for scalable long-context inference on resource-constrained mobile devices and open new opportunities for local-first personal intelligence that increasingly rely on persistent and reusable context.

%Experimental results show substantial prefill-latency improvements over existing on-device LLM serving baselines, demonstrating the strong potential of KV reuse for long-context on-device LLM serving.

\begin{acks}
We sincerely thank all the reviewers and our anonymous shepherd for instructive comments. This work was supported in part by China NSF grant (No. 62572299, No. 62441236, No. 62372296, No. 62432007, No. U24A20326, No. U25A6024, No. U25A20437), the Key Research and Development Program of Zhejiang Province (No. 2024C03270), Alibaba Innovation Research (AIR) Program (No. 56657574-1), CCF-Tencent Rhino-Bird Open Research Fund (No. RAGR20260126), and SJTU-Huawei Explore X Gift Fund.
\end{acks}

% \clearpage

% \balance

\bibliographystyle{ACM-Reference-Format}
\bibliography{ref-eurosys}

@inproceedings{CacheBlend,
author = {Yao, Jiayi and Li, Hanchen and Liu, Yuhan and Ray, Siddhant and Cheng, Yihua and Zhang, Qizheng and Du, Kuntai and Lu, Shan and Jiang, Junchen},
title = {CacheBlend: Fast Large Language Model Serving for RAG with Cached Knowledge Fusion},
year = {2025},
publisher = {Association for Computing Machinery},
address = {New York, NY, USA},
booktitle = {Proceedings of the Twentieth European Conference on Computer Systems},
pages = {94–109},
numpages = {16},
location = {Rotterdam, Netherlands},
series = {EuroSys '25}
}

@misc{LMCache,
      title={LMCache: An Efficient KV Cache Layer for Enterprise-Scale LLM Inference}, 
      author={Yuhan Liu and Yihua Cheng and Jiayi Yao and Yuwei An and Xiaokun Chen and Shaoting Feng and Yuyang Huang and Samuel Shen and Rui Zhang and Kuntai Du and Junchen Jiang},
      year={2025},
      eprint={2510.09665},
      archivePrefix={arXiv},
      primaryClass={cs.LG},
      url={https://arxiv.org/abs/2510.09665}, 
}

@article{RAGCache,
author = {Jin, Chao and Zhang, Zili and Jiang, Xuanlin and Liu, Fangyue and Liu, Shufan and Liu, Xuanzhe and Jin, Xin},
title = {RAGCache: Efficient Knowledge Caching for Retrieval-Augmented Generation},
year = {2025},
issue_date = {February 2026},
publisher = {Association for Computing Machinery},
address = {New York, NY, USA},
volume = {44},
number = {1},
journal = {ACM Trans. Comput. Syst.},
month = nov,
articleno = {2},
numpages = {27},
}

@inproceedings {Strata,
author = {Zhiqiang Xie and Ziyi Xu and Mark Zhao and Yuwei An and Vikram Sharma Mailthody and Scott Mahlke and Michael Garland and Christos Kozyrakis},
title = {Strata: Hierarchical Context Caching for Long Context Language Model Serving},
booktitle = {20th USENIX Symposium on Operating Systems Design and Implementation (OSDI' 26)},
year = {2026},
pages = {1--16},
publisher = {USENIX Association},
month = jul
}

@inproceedings{vllm,
author = {Kwon, Woosuk and Li, Zhuohan and Zhuang, Siyuan and Sheng, Ying and Zheng, Lianmin and Yu, Cody Hao and Gonzalez, Joseph and Zhang, Hao and Stoica, Ion},
title = {Efficient Memory Management for Large Language Model Serving with PagedAttention},
year = {2023},
isbn = {9798400702297},
publisher = {Association for Computing Machinery},
address = {New York, NY, USA},
booktitle = {Proceedings of the 29th Symposium on Operating Systems Principles},
pages = {611–626},
numpages = {16},
series = {SOSP '23}
}

@inproceedings{SGLang,
author = {Zheng, Lianmin and Yin, Liangsheng and Xie, Zhiqiang and Sun, Chuyue and Huang, Jeff and Yu, Cody Hao and Cao, Shiyi and Kozyrakis, Christos and Stoica, Ion and Gonzalez, Joseph E. and Barrett, Clark and Sheng, Ying},
title = {SGLang: efficient execution of structured language model programs},
year = {2024},
publisher = {Curran Associates Inc.},
address = {Red Hook, NY, USA},
booktitle = {Proceedings of the 38th International Conference on Neural Information Processing Systems},
articleno = {2000},
numpages = {27},
location = {Vancouver, BC, Canada},
series = {NIPS '24}
}

@article{Mooncake,
author = {Qin, Ruoyu and Li, Zheming and He, Weiran and Cui, Jialei and Tang, Heyi and Ren, Feng and Ma, Teng and Cai, Shangming and Zhang, Yineng and Zhang, Mingxing and Wu, Yongwei and Zheng, Weimin and Xu, Xinran},
title = {Mooncake: A KVCache-centric Disaggregated Architecture for LLM Serving},
year = {2025},
publisher = {Association for Computing Machinery},
address = {New York, NY, USA},
journal = {ACM Trans. Storage},
month = nov,
}

@inproceedings{PromptCache,
 author = {Gim, In and Chen, Guojun and Lee, Seung-seob and Sarda, Nikhil and Khandelwal, Anurag and Zhong, Lin},
 booktitle = {Proceedings of Machine Learning and Systems},
 pages = {325--338},
 title = {Prompt Cache: Modular Attention Reuse for Low-Latency Inference},
 volume = {6},
 year = {2024},
 series = {MLsys '24}
}

@inproceedings{ChunkAttention,
    title = "{C}hunk{A}ttention: Efficient Self-Attention with Prefix-Aware {KV} Cache and Two-Phase Partition",
    author = "Ye, Lu  and
      Tao, Ze  and
      Huang, Yong  and
      Li, Yang",
    booktitle = "Proceedings of the 62nd Annual Meeting of the Association for Computational Linguistics (Volume 1: Long Papers)",
    year = "2024",
    address = "Bangkok, Thailand",
    publisher = "Association for Computational Linguistics",
    pages = "11608--11620",
    series = {ACL '24}
}

@inproceedings{mllm,
author = {Xu, Daliang and Zhang, Hao and Yang, Liming and Liu, Ruiqi and Huang, Gang and Xu, Mengwei and Liu, Xuanzhe},
title = {Fast On-device LLM Inference with NPUs},
year = {2025},
publisher = {Association for Computing Machinery},
address = {New York, NY, USA},
booktitle = {Proceedings of the 30th ACM International Conference on Architectural Support for Programming Languages and Operating Systems, Volume 1},
pages = {445–462},
numpages = {18},
location = {Rotterdam, Netherlands},
series = {ASPLOS '25}
}

@misc{executorch,
	author = {PyTorch Foundation},
	title = {{G}it{H}ub - pytorch/executorch: {O}n-device {A}{I} across mobile, embedded and edge for {P}y{T}orch},
	howpublished = {\url{https://github.com/pytorch/executorch}},
	year = {2025},
}

@inproceedings {Walle, 
   author = {Chengfei Lv and Chaoyue Niu and Renjie Gu and Xiaotang Jiang and Zhaode Wang and Bin Liu and Ziqi Wu and Qiulin Yao and Congyu Huang and Panos Huang and Tao Huang and Hui Shu and Jinde Song and Bin Zou and Peng Lan and Guohuan Xu and Fei Wu and Shaojie Tang and Fan Wu and Guihai Chen}, 
   title = {Walle: An {End-to-End}, {General-Purpose}, and {Large-Scale} Production System for {Device-Cloud} Collaborative Machine Learning},
   booktitle = {Proceedings of {USENIX} Symposium on Operating Systems Design and Implementation}, 
   year = {2022}, 
   publisher = {USENIX}, 
   address = {Carlsbad, CA, USA},
   pages = {249--265}, 
   series = {OSDI '22}
}

@inproceedings{MNN-LLM,
author = {Wang, Zhaode and Yang, Jingbang and Qian, Xinyu and Xing, Shiwen and Jiang, Xiaotang and Lv, Chengfei and Zhang, Shengyu},
title = {MNN-LLM: A Generic Inference Engine for Fast Large Language Model Deployment on Mobile Devices},
year = {2024},
publisher = {Association for Computing Machinery},
address = {New York, NY, USA},
booktitle = {Proceedings of the 6th ACM International Conference on Multimedia in Asia Workshops},
articleno = {11},
numpages = {7},
series = {MMAsia '24 Workshops}
}

@misc{PowerInfer-2,
      title={PowerInfer-2: Fast Large Language Model Inference on a Smartphone}, 
      author={Zhenliang Xue and Yixin Song and Zeyu Mi and Xinrui Zheng and Yubin Xia and Haibo Chen},
      year={2024},
      eprint={2406.06282},
      archivePrefix={arXiv},
      primaryClass={cs.LG},
      url={https://arxiv.org/abs/2406.06282}, 
}

@misc{llama.cpp,
	author = {Georgi Gerganov},
	title = {{G}it{H}ub - ggml-org/llama.cpp: {L}{L}{M} inference in {C}/{C}++},
	howpublished = {\url{https://github.com/ggml-org/llama.cpp}},
	year = {2026},
}

@misc{T-MAN,
      title={T-MAN: Enabling End-to-End Low-Bit LLM Inference on NPUs via Unified Table Lookup}, 
      author={Jianyu Wei and Qingtao Li and Shijie Cao and Lingxiao Ma and Zixu Hao and Yanyong Zhang and Xiaoyan Hu and Ting Cao},
      year={2025},
      eprint={2511.11248},
      archivePrefix={arXiv},
      primaryClass={cs.AR},
      url={https://arxiv.org/abs/2511.11248}, 
}

@inproceedings{T-MAC,
author = {Wei, Jianyu and Cao, Shijie and Cao, Ting and Ma, Lingxiao and Wang, Lei and Zhang, Yanyong and Yang, Mao},
title = {T-MAC: CPU Renaissance via Table Lookup for Low-Bit LLM Deployment on Edge},
year = {2025},
publisher = {Association for Computing Machinery},
address = {New York, NY, USA},
booktitle = {Proceedings of the Twentieth European Conference on Computer Systems},
pages = {278–292},
numpages = {15},
location = {Rotterdam, Netherlands},
series = {EuroSys '25}
}

@software{MLC-LLM,
    author = {{MLC team}},
    organization = {{CMU Foundation and Language Model Center}},
    title = {{MLC-LLM}},
    url = {https://github.com/mlc-ai/mlc-llm},
    year = {2023-2025}
}

@inproceedings{TVM,
author = {Chen, Tianqi and Moreau, Thierry and Jiang, Ziheng and Zheng, Lianmin and Yan, Eddie and Cowan, Meghan and Shen, Haichen and Wang, Leyuan and Hu, Yuwei and Ceze, Luis and Guestrin, Carlos and Krishnamurthy, Arvind},
title = {TVM: an automated end-to-end optimizing compiler for deep learning},
year = {2018},
isbn = {9781931971478},
publisher = {USENIX Association},
address = {Carlsbad, CA, USA},
booktitle = {Proceedings of the 13th USENIX Conference on Operating Systems Design and Implementation},
pages = {579–594},
numpages = {16},
series = {OSDI '18}
}

@misc{MediaPipe,
	author = {Google AI Edge},
	title = {{M}edia{P}ipe {S}olutions guide},
	howpublished = {\url{https://ai.google.dev/edge/mediapipe/solutions/guide}},
	year = {2025},
}

@inproceedings{llm-npu,
author = {Hao, Zixu and Wei, Jianyu and Wang, Tuowei and Huang, Minxing and Jiang, Huiqiang and Jiang, Shiqi and Cao, Ting and Ren, Ju},
title = {Scaling LLM Test-Time Compute with Mobile NPU on Smartphones},
year = {2026},
isbn = {9798400722127},
publisher = {Association for Computing Machinery},
address = {New York, NY, USA},
booktitle = {Proceedings of the 21st European Conference on Computer Systems},
pages = {2157–2172},
numpages = {16},
series = {Eurosys '26}
}

@inproceedings{HeteroLLM,
author = {Chen, Le and Feng, Dahu and Feng, Erhu and Wang, Yingrui and Zhao, Rong and Xia, Yubin and Xu, Pinjie and Chen, Haibo},
title = {Characterizing Mobile SoC for Accelerating Heterogeneous LLM Inference},
year = {2025},
publisher = {Association for Computing Machinery},
address = {New York, NY, USA},
booktitle = {Proceedings of the ACM SIGOPS 31st Symposium on Operating Systems Principles},
pages = {359–374},
series = {SOSP '25}
}

@misc{LiteRT,
	author = {Google AI Edge},
	title = {{L}iteRT overview},
	howpublished = {\url{https://ai.google.dev/edge/litert}},
	year = {2025},
}

@inproceedings{SmartMem,
author = {Niu, Wei and Sanim, Md Musfiqur Rahman and Shu, Zhihao and Guan, Jiexiong and Shen, Xipeng and Yin, Miao and Agrawal, Gagan and Ren, Bin},
title = {SmartMem: Layout Transformation Elimination and Adaptation for Efficient DNN Execution on Mobile},
year = {2024},
isbn = {9798400703867},
publisher = {Association for Computing Machinery},
address = {New York, NY, USA},
booktitle = {Proceedings of the 29th ACM International Conference on Architectural Support for Programming Languages and Operating Systems, Volume 3},
pages = {916–931},
numpages = {16},
series = {ASPLOS '24}
}

@inproceedings{SpinQuant,
title={SpinQuant: {LLM} Quantization with Learned Rotations},
author={Zechun Liu and Changsheng Zhao and Igor Fedorov and Bilge Soran and Dhruv Choudhary and Raghuraman Krishnamoorthi and Vikas Chandra and Yuandong Tian and Tijmen Blankevoort},
booktitle={Proceedings of The Thirteenth International Conference on Learning Representations (ICLR '25)},
numpages={24},
year={2025},
}

@inproceedings{SmoothQuant,
author = {Xiao, Guangxuan and Lin, Ji and Seznec, Mickael and Wu, Hao and Demouth, Julien and Han, Song},
title = {SmoothQuant: accurate and efficient post-training quantization for large language models},
year = {2023},
publisher = {JMLR.org},
booktitle = {Proceedings of the 40th International Conference on Machine Learning},
articleno = {1585},
numpages = {13},
series = {ICML '23}
}

@inproceedings{MicroNN,
author = {Pound, Jeffrey and Chabert, Floris and Bhushan, Arjun and Goswami, Ankur and Pacaci, Anil and Chowdhury, Shihabur Rahman},
title = {MicroNN: An On-device Disk-resident Updatable Vector Database},
year = {2025},
publisher = {Association for Computing Machinery},
address = {New York, NY, USA},
booktitle = {Companion of the 2025 International Conference on Management of Data},
pages = {608–621},
numpages = {14},
location = {Berlin, Germany},
series = {SIGMOD/PODS '25}
}

@article{SQLite,
author = {Gaffney, Kevin P. and Prammer, Martin and Brasfield, Larry and Hipp, D. Richard and Kennedy, Dan and Patel, Jignesh M.},
title = {SQLite: past, present, and future},
year = {2022},
issue_date = {August 2022},
publisher = {VLDB Endowment},
volume = {15},
number = {12},
journal = {Proc. VLDB Endow.},
month = aug,
pages = {3535–3547},
numpages = {13}
}

@software{FlatBuffers,
    organization = {{Google}},
    title = {FlatBuffers: Memory Efficient Serialization Library},
    url = {https://github.com/google/flatbuffers},
    year = {2025}
}

@inproceedings{SFSL,
author = {Niu, Chaoyue and Wu, Fan and Tang, Shaojie and Hua, Lifeng and Jia, Rongfei and Lv, Chengfei and Wu, Zhihua and Chen, Guihai},
title = {Billion-scale federated learning on mobile clients: a submodel design with tunable privacy},
year = {2020},
publisher = {Association for Computing Machinery},
address = {New York, NY, USA},
booktitle = {Proceedings of the 26th Annual International Conference on Mobile Computing and Networking},
articleno = {31},
numpages = {14},
location = {London, United Kingdom},
series = {MobiCom '20}
}

@ARTICLE{SparKV,
  author={Liu, Hongyao and Zhai, Liuqun and Wang, Junyi and Fang, Zhengru and Chen, Jingshu and Huang, Jun},
  journal={IEEE Internet of Things Journal}, 
  title={SparKV: Overhead-Aware KV Cache Loading for Efficient On-Device LLM Inference}, 
  year={2026},
  volume={13},
  number={14},
  pages={30016-30027},
}

@inproceedings{MobiLoRA,
    title = "{M}obi{L}o{RA}: Accelerating {L}o{RA}-based {LLM} Inference on Mobile Devices via Context-aware {KV} Cache Optimization",
    author = "Li, Borui  and
      Wang, Yitao  and
      Ma, Haoran  and
      Chen, Ligeng  and
      Xiao, Jun  and
      Wang, Shuai",
    booktitle = "Proceedings of the 63rd Annual Meeting of the Association for Computational Linguistics (Volume 1: Long Papers)",
    month = jul,
    year = "2025",
    address = "Vienna, Austria",
    publisher = "Association for Computational Linguistics",
    pages = "23400--23410",
}

@inproceedings{InfiniGen,
author = {Lee, Wonbeom and Lee, Jungi and Seo, Junghwan and Sim, Jaewoong},
title = {InfiniGen: efficient generative inference of large language models with dynamic KV cache management},
year = {2024},
publisher = {USENIX Association},
address = {USA},
booktitle = {Proceedings of the 18th USENIX Conference on Operating Systems Design and Implementation},
articleno = {9},
numpages = {18},
location = {Santa Clara, CA, USA},
series = {OSDI'24}
}

@misc{Llama3,
      title={The Llama 3 Herd of Models}, 
      author={Llama Team},
      year={2024},
      eprint={2407.21783},
      archivePrefix={arXiv},
      primaryClass={cs.AI},
      url={https://arxiv.org/abs/2407.21783}, 
}

@misc{Llama3.2,
	author = {Llama Team},
	title = {{L}lama 3.2: {R}evolutionizing edge {A}{I} and vision with open, customizable models},
	howpublished = {\url{https://ai.meta.com/blog/llama-3-2-connect-2024-vision-edge-mobile-devices/}},
	year = {2024},
}

@misc{Qwen3,
  author={Qwen Team},
      title={Qwen3 Technical Report}, 
      year={2025},
      eprint={2505.09388},
      archivePrefix={arXiv},
      primaryClass={cs.CL},
      url={https://arxiv.org/abs/2505.09388}, 
}

@misc{Qwen3-Embedding,
      title={Qwen3 Embedding: Advancing Text Embedding and Reranking Through Foundation Models}, 
      author={Yanzhao Zhang and Mingxin Li and Dingkun Long and Xin Zhang and Huan Lin and Baosong Yang and Pengjun Xie and An Yang and Dayiheng Liu and Junyang Lin and Fei Huang and Jingren Zhou},
      year={2025},
      eprint={2506.05176},
      archivePrefix={arXiv},
      primaryClass={cs.CL},
      url={https://arxiv.org/abs/2506.05176}, 
}

@misc{Apple-Intelligence-Device-Cloud,
	author = {Apple},
	title = {{I}ntroducing {A}pple’s {O}n-{D}evice and {S}erver {F}oundation {M}odels},
	howpublished = {\url{https://machinelearning.apple.com/research/introducing-apple-foundation-models}},
	year = {2024}
}

@inproceedings{ClawMobile,
author = {Du, Hongchao and Wu, Shangyu and Li, Qiao and Pan, Riwei and Li, Jinheng and Sun, Youcheng and Xue, Chun Jason},
title = {ClawMobile: Rethinking Smartphone-Native Agentic Systems},
year = {2026},
publisher = {Association for Computing Machinery},
booktitle = {Proceedings of the Sixth European Workshop on Machine Learning and Systems},
pages = {370–376},
numpages = {7},
series = {EuroMLSys '26}
}

@misc{MobileRAG,
      title={MobileRAG: A Fast, Memory-Efficient, and Energy-Efficient Method for On-Device RAG}, 
      author={Taehwan Park and Geonho Lee and Min-Soo Kim},
      year={2025},
      eprint={2507.01079},
      archivePrefix={arXiv},
      primaryClass={cs.DB},
      url={https://arxiv.org/abs/2507.01079}, 
}

@misc{OmniInfer,
      title={OmniInfer: System-Wide Acceleration Techniques for Optimizing LLM Serving Throughput and Latency}, 
      author={Jun Wang and Yunxiang Yao and Wenwei Kuang and Runze Mao and Zhenhao Sun and Zhuang Tao and Ziyang Zhang and Dengyu Li and Jiajun Chen and Zhili Wang and Kai Cui and Congzhi Cai and Longwen Lan and Ken Zhang},
      year={2025},
      eprint={2511.22481},
      archivePrefix={arXiv},
      primaryClass={cs.DC},
      url={https://arxiv.org/abs/2511.22481}, 
}

@inproceedings{HotpotQA,
    title = "{H}otpot{QA}: A Dataset for Diverse, Explainable Multi-hop Question Answering",
    author = "Yang, Zhilin  and
      Qi, Peng  and
      Zhang, Saizheng  and
      Bengio, Yoshua  and
      Cohen, William  and
      Salakhutdinov, Ruslan  and
      Manning, Christopher D.",
    booktitle = "Proceedings of the 2018 Conference on Empirical Methods in Natural Language Processing",
    month = oct # "-" # nov,
    year = "2018",
    address = "Brussels, Belgium",
    publisher = "Association for Computational Linguistics",
    pages = "2369--2380",
}

@inproceedings{LoCoMo,
    title = "Evaluating Very Long-Term Conversational Memory of {LLM} Agents",
    author = "Maharana, Adyasha  and
      Lee, Dong-Ho  and
      Tulyakov, Sergey  and
      Bansal, Mohit  and
      Barbieri, Francesco  and
      Fang, Yuwei",
    booktitle = "Proceedings of the 62nd Annual Meeting of the Association for Computational Linguistics (Volume 1: Long Papers)",
    month = aug,
    year = "2024",
    address = "Bangkok, Thailand",
    publisher = "Association for Computational Linguistics",
    pages = "13851--13870",
}

@misc{LoCoMo‑MC10,
      title={LoCoMo‑MC10 · Long Conversation Memory Multiple‑Choice 10}, 
      author={Percena},
      year={2025},
      url={https://huggingface.co/datasets/Percena/locomo-mc10}, 
}

@misc{SkillsBench,
	author = {benchflow-ai},
	title = {SkillsBench:The first benchmark for evaluating how well AI agents use skills.},
	howpublished = {\url{https://github.com/benchflow-ai/skillsbench}},
	year = {2026},
}

@inproceedings{LLM-in-a-Flash,
    title = "{LLM} in a flash: Efficient Large Language Model Inference with Limited Memory",
    author = "Alizadeh, Keivan  and
      Mirzadeh, Seyed Iman  and
      Belenko, Dmitry  and
      Khatamifard, S.  and
      Cho, Minsik  and
      Del Mundo, Carlo C  and
      Rastegari, Mohammad  and
      Farajtabar, Mehrdad",
    booktitle = "Proceedings of the 62nd Annual Meeting of the Association for Computational Linguistics (Volume 1: Long Papers)",
    year = "2024",
    address = "Bangkok, Thailand",
    publisher = "Association for Computational Linguistics",
    pages = "12562--12584",
}

@inproceedings{Neuralink,
author = {Wang, Tuowei and Fan, Ruwen and Huang, Minxing and Hao, Zixu and Li, Kun and Cao, Ting and Lu, Youyou and Zhang, Yaoxue and Ren, Ju},
title = {{Neuralink}: Fast on-Device LLM Inference with Neuron Co-Activation Linking},
year = {2025},
isbn = {9798400710803},
publisher = {Association for Computing Machinery},
address = {New York, NY, USA},
booktitle = {Proceedings of the 30th ACM International Conference on Architectural Support for Programming Languages and Operating Systems, Volume 3},
pages = {147–162},
numpages = {16},
series = {ASPLOS '25}
}

@misc{Apple-Intelligence,
	author = {{Apple Inc.}},
	title = {{A}pple {I}ntelligence},
	howpublished = {\url{https://www.apple.com/apple-intelligence/}},
	year = {2026},
}

@software{GoogleAI,
  author = {{Google}},
  title  = {{Google AI Edge Gallery}},
  year   = {2026},
  url    = {https://play.google.com/store/apps/details?id=com.google.ai.edge.gallery},
}

@online{google2026gemini,
  title        = {Gemini Introduces Personal Intelligence},
  year         = {2026},
  month        = jan,
  day          = {14},
  organization = {Google},
  url          = {https://blog.google/innovation-and-ai/products/gemini-app/personal-intelligence/},
  urldate      = {2026-05-05},
}

@misc{hexagon-htp,
	author = {Qualcomm Technologies Inc.},
	title = {{Q}ualcomm {D}ocumentation --- {HTP} {Backend}},
	howpublished = {\url{https://docs.qualcomm.com/doc/80-63442-10/topic/htp_backend.html}},
	year = {2026},
}

@misc{NeuroPilot,
    author = {MediaTek Inc.},
    title = {NeuroPilot Documentation},
    howpublished = {\url{https://neuropilot-developer.mediatek.com/sphinx/neuropilot-8-public/html/}},
    year = {2026},
}

@misc{CoreML,
	author = {Apple Inc.},
	title = {CoreML Documentation},
	howpublished = {\url{https://developer.apple.com/documentation/coreml}},
	year = {2026},
}

% \clearpage

\nobalance

\appendix

\section{Graph Structure and Quantization Scheme of NPU Selective Recompute Graph}
\begin{figure}[ht]
    \centering
    \includegraphics[width=\linewidth]{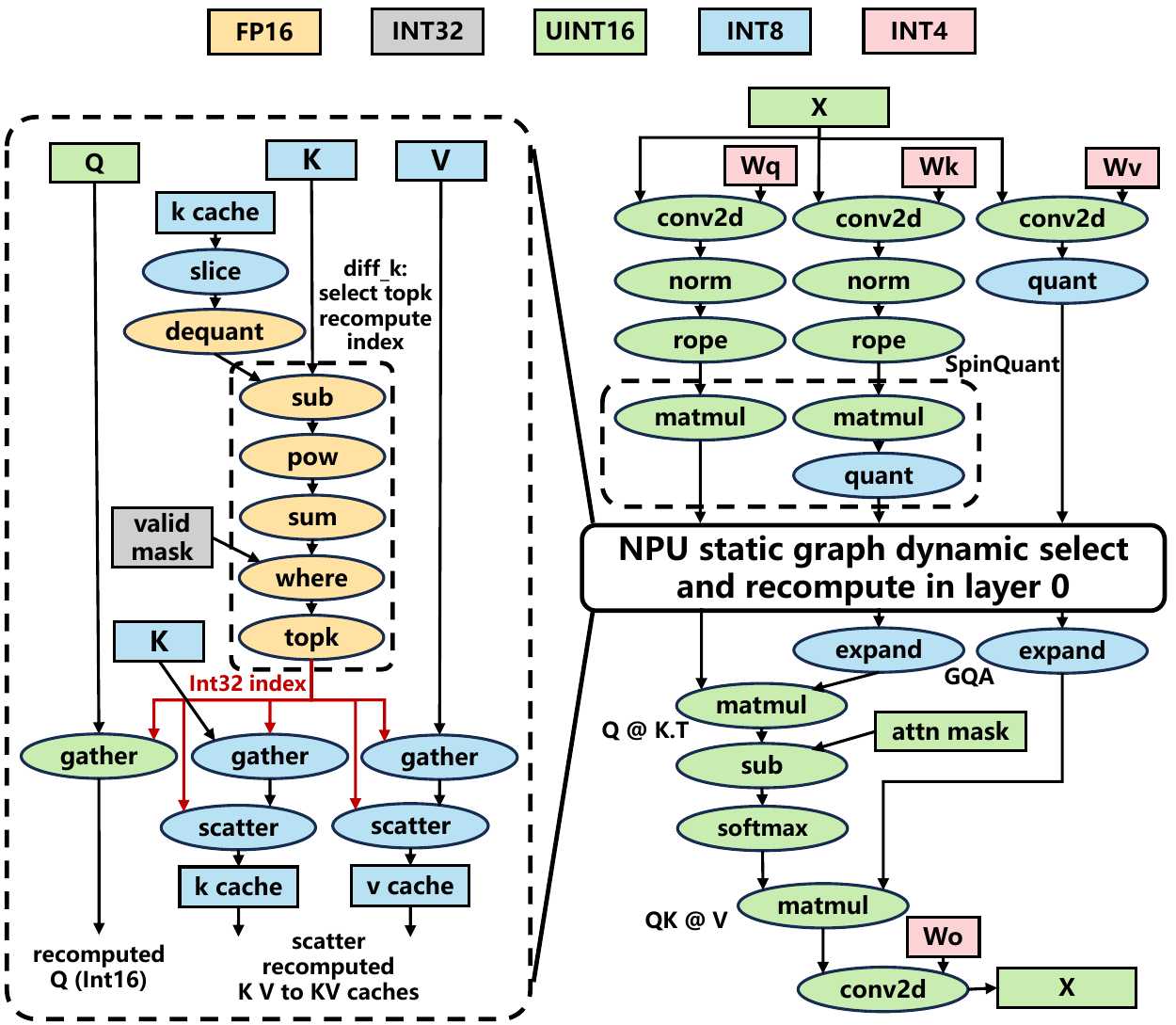}
    \caption{Quantized NPU Selective KV Recompute Graph of Layer 0.}
    \Description{Quantized NPU Selective KV Recompute Graph of Layer 0.}
    \label{fig:quant-graph}
\end{figure}

We follow the $W_{\mathrm{INT4}}A_{\mathrm{UINT16}}KV_{\mathrm{INT8}}$ quantization  adopted in ExecuTorch~\cite{ExecuTorch}. 
Specifically, KV tensors are quantized to 8-bit using SpinQuant~\cite{SpinQuant}, which applies a Hadamard transformation to mitigate the accuracy loss caused by severe outliers in KV distributions~\cite{SpinQuant, SmoothQuant}. 
On the other hand, the precision-critical KV deviation computation and Top-K selection are retained in FP16 precision, while KV tensors are SpinQuanted to INT8, and most intermediate activations use UINT16. 
The graph structure and corresponding bit-width is illustrated in Figure~\ref{fig:quant-graph}.

\section{Chunk Merge Algorithm for Efficient Static Graph Utilization}
\label{app:chunk_merge}

  We give the complete formulation of the chunk merge dynamic program.

  \paragraph{Setup.}
  Let the token sequence be partitioned into ordered segments
  \[
  P_j=(s_j,e_j,p_j), \qquad j=0,\dots,m-1,
  \]
  where $[s_j,e_j]$ is the span of segment $j$ and
  $p_j\in\{0,1,2\}$ indicates prefix reuse, non-prefix reuse, and new tokens, respectively.
  Let $L_s$ and $L_p$ be the token capacity of the selective recomputation graph and prefill graph, and let $t_s$
  and $t_p$ be their latency.
  Let $r\in(0,1]$ be the minimum recomputation ratio.

  We define $T[i]$ as the minimum latency to process tokens up to position $i$.
  The base case is that prefix reuse tokens require no computation, so if position $i$ lies entirely in the prefix-
  reuse region, then
  $T[i]=0$.

  \paragraph{Transition.}
  For a chunk ending at token position $i$, we consider two options:
  \begin{equation}
      T[i] = \min \bigl\{ T[i-l_s(i)] + t_s,\; T[i-l_p(i)] + t_p \bigr\}.
  \end{equation}
  Here:
  \begin{itemize}
  \item $l_s(i)$ is the longest valid suffix ending at $i$ that can be packed into one selective recomputation call.
  \item $l_p(i)$ is the longest valid suffix ending at $i$ that can be packed into one prefill call.
  \end{itemize}

  The algorithmic implementation revolves around this transition function, and the pseudo code is presented in Algorithm \ref{alg:chunk_merge}. 

  \begin{algorithm}[h]
  \caption{Dynamic Programming Chunk Merge}
  \label{alg:chunk_merge}
  \KwIn{
  Segment list $P_j=(s_j,e_j,p_j)$ for $j=0,\dots,m-1$; \\
  selective recomputation capacity $L_s$ and latency $t_s$; \\
  prefill capacity $L_p$ and latency $t_p$; \\
  minimum recomputation ratio $r$
  }
  \KwOut{
  Minimum latency $T[L-1]$ and chunk decisions
  }

  $L \gets e_{m-1}+1$\;
  Initialize $T[i]\gets +\infty$ and $\mathrm{dec}[i]\gets \varnothing$ for all $i=0,\dots,L-1$\;

  \For{$i\gets 0$ \KwTo $L-1$}{
      \If{$i$ is in the prefix reuse region}{
          $T[i]\gets 0$\;
          continue\;
      }

      $l_s \gets \mathrm{ComputeSRLength}(P, i, L_s, r)$\;
      $l_p \gets \mathrm{ComputePrefillLength}(P, i, L_p)$\;

      $v_s \gets T[i-l_s] + t_s$ \tcp*{use $0$ if $i-l_s<0$}
      $v_p \gets T[i-l_p] + t_p$ \tcp*{use $0$ if $i-l_p<0$}

      \eIf{$v_s < v_p$}{
          $T[i]\gets v_s$\;
          $\mathrm{dec}[i]\gets (i-l_s,\mathrm{SR})$\;
      }{
          $T[i]\gets v_p$\;
          $\mathrm{dec}[i]\gets (i-l_p,\mathrm{P})$\;
      }
  }

  \Return $T[L-1]$ and $\mathrm{dec}$\;
  \end{algorithm}

  \paragraph{Computation of $l_p(i)$.}
  The prefill graph simply packs as many trailing tokens as allowed by its capacity.
  Starting from token $i$ and scanning backward over segments, we accumulate tokens until either:
  (1) the prefill budget $L_p$ is exhausted, or
  (2) we reach the prefix reuse region.
  Thus, if the accumulated packed length is $\ell$, then $l_p(i)=\ell$.

  \paragraph{Computation of $l_s(i)$.}
  The selective recomputation graph has capacity $L_s$, but new tokens may only occupy a bounded portion of that
  capacity.
  When scanning backward from token $i$:
  \begin{itemize}
  \item if the current segment is non-prefix reuse ($p_j=1$), its tokens consume recomputation capacity directly;
  \item if the current segment is new ($p_j=2$), only a bounded number of its tokens may be merged into the current
  recomputation call so that the recomputation ratio remains at least $r$.
  \end{itemize}
  Equivalently, if the remaining recomputation capacity is $q$, then at most $\lfloor q r \rfloor$ new tokens may be
  absorbed from the current new-token segment.
  If $\Delta$ new tokens are absorbed, they consume $\lceil \Delta / r \rceil$ units of recomputation capacity.
  Scanning stops when the recomputation capacity is exhausted or when the prefix reuse region is reached.
  If the total absorbed suffix length is $\ell$, then $l_s(i)=\ell$.

The functions for computing $l_p(i)$ and $l_s(i)$ are presented in Algorithm \ref{alg:compute_lp} and \ref{alg:compute_ls} respectively.

  % \paragraph{Optimality.}
  % The recurrence is optimal because the last graph invocation in an optimal schedule must be either a selective
  % recomputation call or a prefill call.
  % Once that final chunk is fixed, the remaining prefix is an independent subproblem of the same form.
  % Hence the problem admits the above optimal substructure.

  \begin{algorithm}[t]
  \caption{$\mathrm{ComputePrefillLength}(P, i, L_p)$}
  \label{alg:compute_lp}
  \KwIn{Segment list $P$, ending position $i$, prefill budget $L_p$}
  \KwOut{$l_p(i)$}

  Locate the segment index $j$ such that $i\in [s_j,e_j]$\;
  $q \gets L_p$, $\ell \gets 0$, $x \gets i$\;

  \While{$q>0$ and segment $j$ is not prefix reuse}{
      $a \gets x-s_j+1$\;
      $\Delta \gets \min(q, a)$\;
      $\ell \gets \ell + \Delta$\;
      $q \gets q - \Delta$\;
      $x \gets x - \Delta$\;
      \If{$x < s_j$}{
          $j \gets j-1$\;
      }
  }

  \Return $\ell$\;
  \end{algorithm}

  \begin{algorithm}[t]
  \caption{$\mathrm{ComputeSRLength}(P, i, L_s, r)$}
  \label{alg:compute_ls}
  \KwIn{Segment list $P$, ending position $i$, recomputation budget $L_s$, ratio $r$}
  \KwOut{$l_s(i)$}

  Locate the segment index $j$ such that $i\in [s_j,e_j]$\;
  $q \gets L_s$, $\ell \gets 0$, $x \gets i$\;

  \While{$q>0$ and segment $j$ is not prefix reuse}{
      $a \gets x-s_j+1$ \tcp*{available suffix length in current segment}

      \uIf{$p_j = 2$}{
          $\Delta \gets \min(\lfloor q r \rfloor, a)$\;
          \If{$\Delta = 0$}{
              \textbf{break}\;
          }
          $\ell \gets \ell + \Delta$\;
          $q \gets q - \lceil \Delta / r \rceil$\;
      }
      \Else{
          $\Delta \gets \min(q, a)$\;
          $\ell \gets \ell + \Delta$\;
          $q \gets q - \Delta$\;
      }

      $x \gets x - \Delta$\;
      \If{$x < s_j$}{
          $j \gets j-1$\;
      }
  }

  \Return $\ell$\;
  \end{algorithm}

\section{Fast LCS for Non-prefix Matching}
To match the reusable substrings in candidate cached chunks, we apply a linear-time LCS (Longest Common Substring) for fast matching. 

To compute the longest common substring between two token sequences, we use a suffix automaton with sparse transitions. 
Given two sequences $X$ and $Y$, where each sequence contains at most 512 tokens (database chunk size limit) and the token vocabulary can be as large as $10^6$, we build the suffix automaton over the shorter sequence $X$ and then stream the longer sequence $Y$ through the automaton. 
During the scan, we maintain the current automaton state and the length of the current match; when the next token transition is absent, we follow suffix links until a valid transition is found or the root is reached. 
The maximum matched length observed during this scan is the longest common token substring length.

The key design choice is to avoid alphabet-dense transition tables. 
Although the global vocabulary is large, each sequence contains only a small number of tokens, and a suffix automaton over $X$ contains at most $2|X|-1$ states and $O(|X|)$ transitions. 
Therefore, we store outgoing transitions sparsely as token-id-to-state mappings, making the memory usage independent of the global vocabulary size. 
With hash-table transitions, transition lookup takes expected $O(1)$ time, so the algorithm runs in expected $O(|X|+|Y|)$ time and uses $O(|X|)$ space. 
For a deterministic worst-case implementation, hash tables can be replaced by sorted sparse transition arrays or balanced maps, giving $O((|X|+|Y|)\log d_{\max})$ time and $O(|X|)$ space, where $d_{\max}\le |X|$ is the maximum outgoing degree of any automaton state. 
Since $|X|,|Y|\le512$, the automaton has at most 1023 states and $\log d_{\max}\le9$, so the deterministic implementation remains approximately linear in practice. The pseudo code is presented in Algorithm \ref{alg:lcs}.
% While advanced deterministic integer dictionaries could in principle provide worst-case $O(|X|+|Y|)$ time with sparse transitions, they introduce unnecessary implementation complexity for our short token sequences.

\begin{algorithm}[h]
\caption{Longest Common Token Substring}
\label{alg:lcs}
\KwIn{Token sequences $A$ and $B$}
\KwOut{Longest common token substring length}

\If{$|A|>|B|$}{
    swap $A$ and $B$\;
}

Build a suffix automaton $\mathcal{S}$ over $A$ with sparse token transitions\;
$v \leftarrow \mathcal{S}.\mathrm{root}$; $\ell \leftarrow 0$; $\mathrm{best}\leftarrow0$\;

\ForEach{token $x \in B$}{
    \While{$v\neq \mathcal{S}.\mathrm{root}$ and $x\notin \mathcal{S}.\mathrm{next}[v]$}{
        $v\leftarrow \mathcal{S}.\mathrm{link}[v]$\;
        $\ell\leftarrow \mathcal{S}.\mathrm{len}[v]$\;
    }

    \eIf{$x\in \mathcal{S}.\mathrm{next}[v]$}{
        $v\leftarrow \mathcal{S}.\mathrm{next}[v][x]$\;
        $\ell\leftarrow \ell+1$\;
    }{
        $v\leftarrow \mathcal{S}.\mathrm{root}$\;
        $\ell\leftarrow0$\;
    }

    $\mathrm{best}\leftarrow\max(\mathrm{best},\ell)$\;
}

\Return $\mathrm{best}$\;
\end{algorithm}

\section{Flash-Storage SQLite-Based DB Organization} 
Directly storing large KV tensors as SQLite BLOB objects can lead to severe internal and external fragmentation, especially under frequent insertion, eviction, and update operations. 
Fragmentation further degrades insertion efficiency because free pages become sparsely scattered, making it difficult for SQLite to allocate sufficiently large contiguous regions for KV BLOBs. 
Moreover, since SQLite page management itself is already built on top of the underlying file system, using SQLite to manage large KV tensors introduces another unnecessary storage-management layer and additional overhead. 
In practice, both insertion and loading latency can increase to the order of seconds if KV storage is handled entirely by SQLite.

\begin{figure}[!h]
    \centering
    \includegraphics[width=0.9\linewidth]{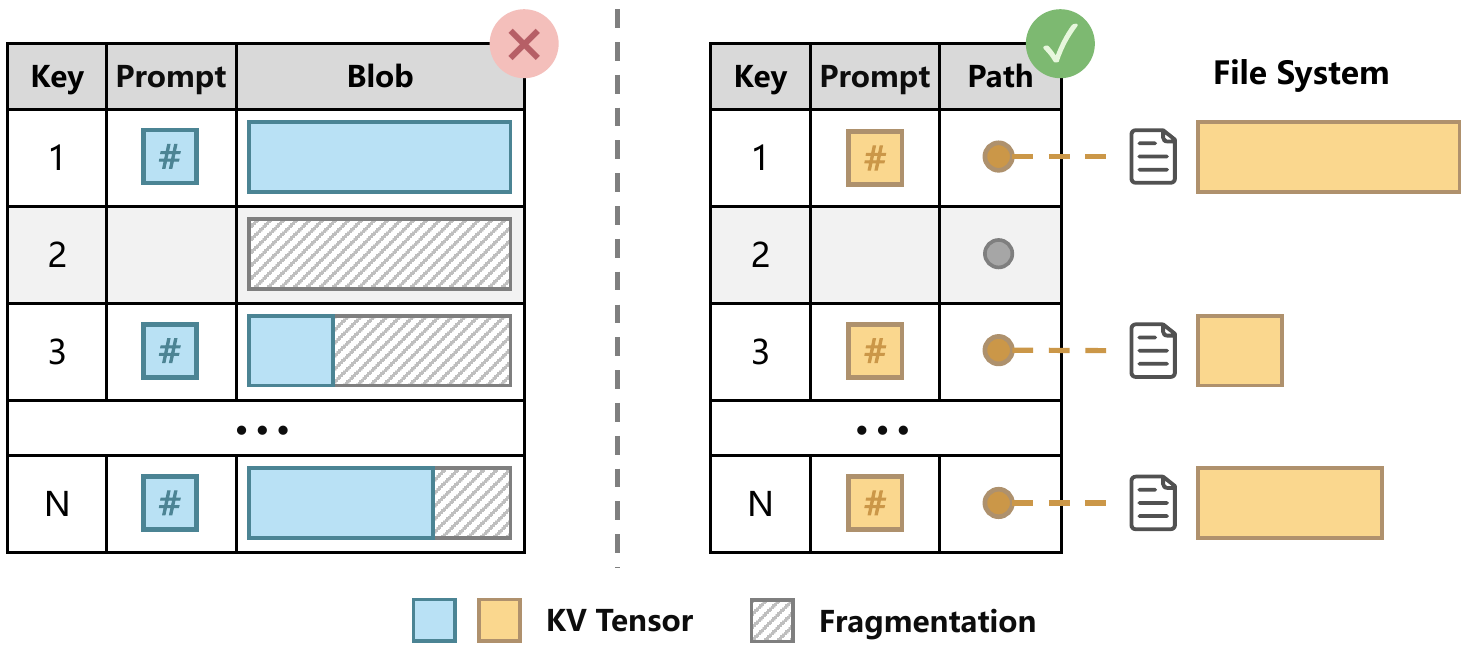}
    \caption{KV tensors blob storage design (self-managed outside SQLite).}
    \label{fig:blob-design}
\end{figure}

Therefore, as illustrated in Figure~\ref{fig:blob-design}, instead of embedding KV tensors directly inside SQLite rows as large BLOB objects, we organize flash storage using a lightweight SQLite metadata index combined with external file-system storage.
Specifically, SQLite stores only compact metadata entries, including chunk keys, prompt hashes, locality statistics, and file paths, while the actual KV tensors are serialized as separate files in the underlying file system. 
This design avoids repeated large-BLOB reallocations within SQLite pages, substantially reducing fragmentation and write amplification. 
It also improves eviction and compaction efficiency, since removing a KV chunk only requires deleting its metadata entry and corresponding file without triggering large-scale database-page reorganization. 
As a result, the storage system maintains stable lookup efficiency while supporting scalable GB-level KV persistence on resource-constrained mobile flash storage.

\section{Locality Model for Prefetch and Reuse}
The locality model estimates the probability that a KV chunk $c_i$ will be reused in the near future. 
The resulting locality score is used to guide both prefetch and eviction decisions across the memory hierarchy.

Conventional cache policies such as LRU and LFU only capture temporal locality, while prefix-aware methods such as PGDSF in RAGCache~\cite{RAGCache} mainly target prefix reuse and cannot effectively model non-prefix reuse behaviors and their associated recomputation costs. 
To address this limitation, we propose a lightweight locality model tailored for mobile long-context workloads, together with an online eviction-cost estimator that jointly considers both prefix and non-prefix reuse.

For a cached chunk $c_i$, the overall locality score is defined as:
\begin{equation}
    R(c_i) = w_sR_s(c_i) + w_tR_t(c_i)+w_mR_m(c_i),
    \label{eq:locality}
\end{equation}
where $R_s$, $R_t$, and $R_m$ denote spatial, temporal, and semantic locality scores, respectively, and $R_s, R_t, R_m$ are normalization weights.

The three locality dimensions are defined as follows:

\begin{itemize}
    \item \emph{Spatial locality} $R_s(c_i)$ captures whether a chunk belongs to the same prefix/non-prefix tree or document as recently reused chunks. 
    This is motivated by the observation that recently accessed documents and histories are more likely to be revisited in the near future.
    
    \item \emph{Temporal locality} $R_t(c_i)$ models both recency and access frequency through a lightweight combination of LRU and LFU signals, allowing frequently accessed chunks to remain in higher memory levels while gradually evicting colder entries.
    
    \item \emph{Semantic locality} $R_m(c_i)$ models correlations among reusable chunks using a lightweight co-access lookup table. 
    Chunks that frequently co-occur in prompts are assigned higher semantic locality scores. 
    For example, SMS-related skills are often co-accessed with website-login skills for verification-code handling, and therefore exhibit strong semantic locality.
\end{itemize}

The eviction policy considers all three locality dimensions, while the prefetch policy only uses spatial and semantic locality to avoid unnecessary memory pollution.

\section{Recomputation/Reloading Cost Model}
The eviction cost model estimates the future overhead introduced by removing a KV chunk from different storage levels. 
Since chunks stored in the CPU memory pool and flash storage incur fundamentally different recovery costs, we model them separately.

\paragraph{CPU memory-pool eviction cost}
For a chunk $c_i$ stored in the CPU memory pool, the eviction cost mainly corresponds to the future reload latency from flash storage:
\begin{equation}
C_{\mathrm{cpu}}(c_i)=
\left\{
\begin{array}{ll}
T_{\mathrm{load}}(c_i), & c_i \in \mathcal{P} \\
0, & c_i \in \mathcal{N}
\end{array}
\right. ,
\label{eq:eviction-CPU}
\end{equation}
where $\mathcal{P}$ and $\mathcal{N}$ denote the sets of prefix-reusable and non-prefix-reusable chunks, respectively, and $T_{\mathrm{load}}(c_i)$ is the flash to memory loading latency of chunk $c_i$. 
For non-prefix-reusable chunks, the loading latency can largely be hidden through compute--storage overlap, making the effective eviction cost close to zero.

\paragraph{Flash-storage eviction cost}
The eviction cost of a chunk stored in flash involves 2 terms: the first is the recomputation cost of the evicted chunk itself, while the second is the additional overhead derived from the cascading effect of breaking the prefix reuse chain. 

Specifically, the primary component is the recomputation cost $Latency_P(|c_i|)$, denoting the full prefill latency of the chunk estimated in Section 4.1.

\begin{figure}[h]
    \centering
    \includegraphics[width=0.63\linewidth]{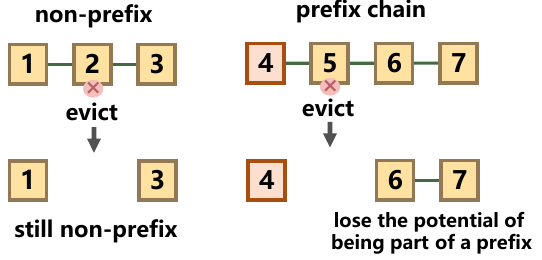}
    \caption{Evicting a chunk on prefix chain makes subsequent chunks lose the potential of being part of a prefix.}
    \label{fig:eviction-cost}
\end{figure}
Besides recomputation cost of the evicted chunk itself, evicting a chunk on a prefix path may cause its descendant chunks to lose prefix reusability and fall back to non-prefix reuse and bring additional cascading cost $C_{\mathrm{cascade}}$. 
As illustrated in Figure~\ref{fig:eviction-cost}, chunks 6 and 7 can originally be reused as part of the prefix chain 4567. 
However, evicting chunk 5 breaks the prefix chain and turns chunks 6 and 7 into pure non-prefix chunks, thereby increasing future recomputation overhead. 
In contrast, evicting chunk 2 introduces no additional cost to chunk 3 because it is already non-prefix-reusable. 
The cascading cost is formulated as:
\begin{equation}
    C_{\mathrm{cascade}} = \sum_{c_j\in \mathrm{Des}(c_i)}f(c_j)\cdot Latency_S(|c_i|),
    \label{eq:cascading-cost}
\end{equation}
where $\mathrm{Des}(c_i)$ denotes the descendant chunks of $c_i$ on the prefix tree, $f(c_j)$ is the prefix-hit frequency factor estimating the likelihood of future prefix reuse, $Latency_S(|c_i|)$ is the additional selective recomputation graph execution overhead caused by turning from prefix to non-prefix.

Combining the 2 terms, the full eviction cost formulate as:
\begin{equation}
C_{\mathrm{flash}}(c_i)=Latency_P(|c_i|)+C_{\mathrm{cascade}}(c_i).
\label{eq:eviction-flash}
\end{equation}

\section{Cost-Aware Prefetch and Eviction}

\subsection{Prefetch}
The prefetch policy proactively loads reusable chunks whose locality scores $R(c)$  greater or equal to a threshold from flash storage into memory to reduce future loading latency. 
Prefetching is performed asynchronously during idle I/O periods and primarily targets prefix-reusable chunks with high predicted reuse probability.

\subsection{Eviction}
The eviction policy ranks chunks according to the locality-cost product: $R(c)\cdot C(c)$.
Chunks with high locality and high recovery cost are preferentially retained in higher memory levels, while less useful chunks are demoted or removed. 
This policy jointly optimizes cache hit rate and recovery overhead under constrained on-device memory and storage capacity.

\end{document}